%% file: ctmva-for-arxiv.tex
\documentclass[12pt]{article}
\usepackage{amsmath, amsfonts, graphicx, commath, natbib, url}
\usepackage[hidelinks]{hyperref}

\input{bold.tex}

\newcommand{\argmax}{\operatornamewithlimits{arg\,max}}

\newcommand{\tr}{\mbox{tr}}

\newcommand{\blind}{0}

\begin{document}

\def\spacingset#1{\renewcommand{\baselinestretch}%
{#1}\small\normalsize} \spacingset{1}


\if0\blind
{
  \title{\bf Continuous-time multivariate analysis}
  \author{Biplab Paul, Philip T. Reiss\thanks{
		The authors thank Hadar Fisher, Ciprian Crainiceanu, Prince Osei, Laura Sangalli and Dror Arbiv for helpful discussions, and the Editor, Associate Editor and referees for valuable feedback. The work of Paul, Reiss and Fo\`{a} was supported by Israel Science Foundation under Grant 1076/19.}\hspace{.2cm}\\
	Department of Statistics, University of Haifa;\\
	Erjia Cui \\
	Division of Biostatistics and Health Data Science, University of Minnesota;\\
	and \\
	Noemi   Fo\`{a}\\
	Department of Statistics, University of Haifa}
  \maketitle
} \fi

\if1\blind
{
\bigskip
\bigskip
\bigskip
\begin{center}
	{\LARGE\bf Continuous-time multivariate analysis}
\end{center}
\medskip
} \fi

\bigskip
\begin{abstract}
The starting point for much of multivariate analysis (MVA) is an $n\times p$ data matrix whose $n$ rows represent observations and whose $p$ columns represent variables. Some multivariate data sets, however, may be best conceptualized not as $n$ discrete $p$-variate observations, but as $p$ curves or functions defined on a common time interval. Here we introduce a framework for extending techniques of multivariate analysis to such settings. The proposed continuous-time multivariate analysis (CTMVA) framework rests on the assumption that the curves can be represented as linear combinations of basis functions such as $B$-splines, as in the Ramsay-Silverman representation of functional data; but whereas functional data analysis extends MVA to the case of observations that are curves rather than vectors --- heuristically, $n\times p$ data with $p$ infinite --- we are instead concerned with what happens when $n$ is infinite. We present continuous-time extensions of the classical MVA methods of covariance and correlation estimation, principal component analysis, Fisher's linear discriminant analysis, and $k$-means clustering. We show that CTMVA can improve on the performance of classical MVA, in particular for correlation estimation and clustering, and  can be applied in some settings where classical MVA cannot, including variables observed at disparate time points. CTMVA is illustrated with a novel perspective on a well-known Canadian weather data set, and with applications to data sets involving international development, brain signals, and air quality. The proposed methods are implemented in the publicly available R package \texttt{ctmva}.
\end{abstract}

\noindent%
\vfill

\newpage
\spacingset{1.5} 

\section{Introduction}
\label{sec:intro}

The generic mathematical object that underpins much of multivariate analysis (MVA) is an $n\times p$ data matrix $\bX$ whose $n$ rows represent observations and whose $p$ columns represent variables. While classical MVA typically assumes the $n$ observations to be independent, there is a long history of applications of MVA to multivariate time series, notably in econometrics and atmospheric science \citep{jolliffe2002}. 
In this paper we argue that some time-indexed multivariate data sets may be best conceptualized not as a time series of $n$ discrete $p$-variate observations, but as $p$ curves or functions defined on a common time interval; and we introduce a framework for straightforwardly extending several techniques of MVA to this setting.

\begin{figure}
	\centering
	\includegraphics[width=.3\textwidth]{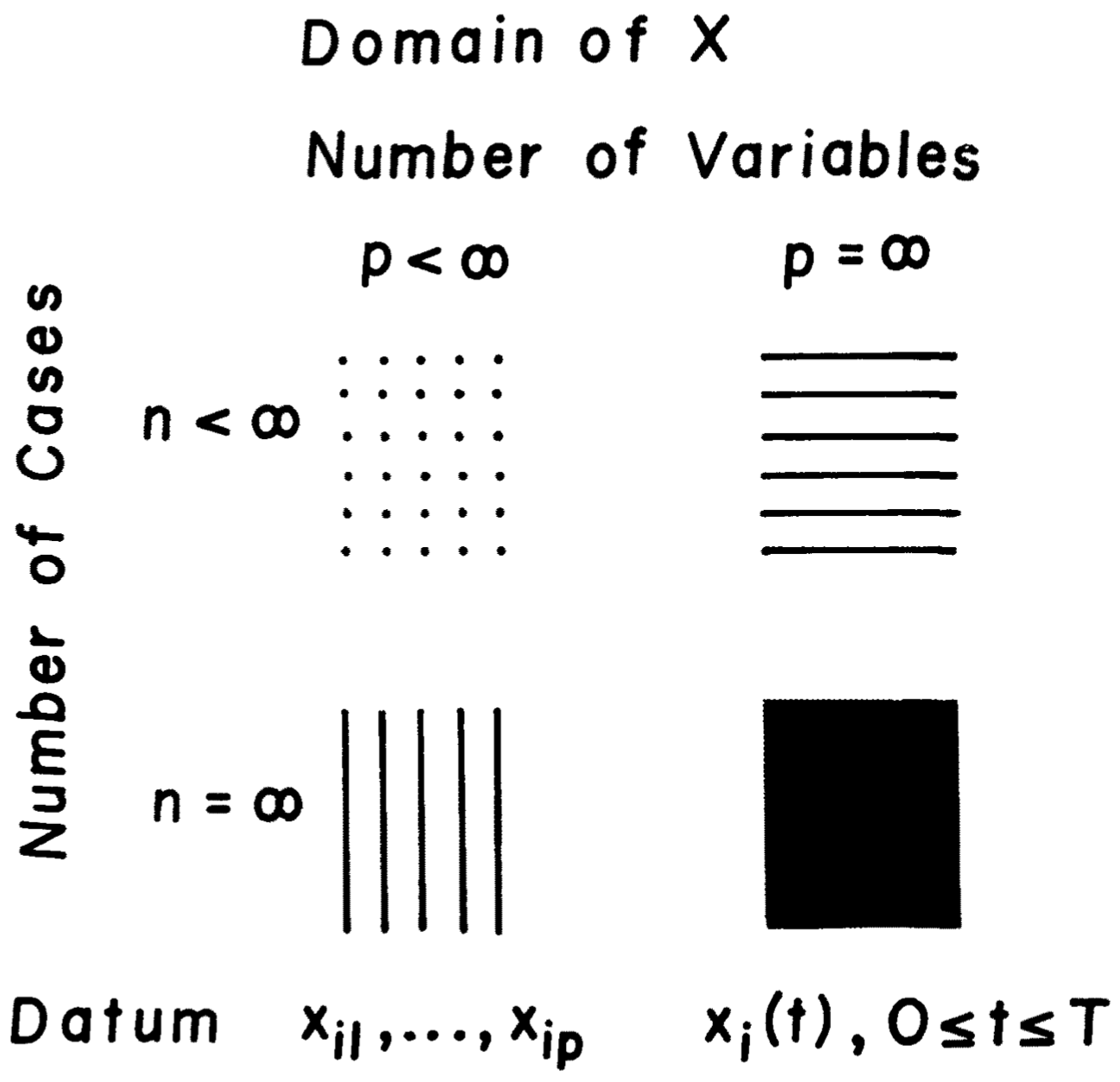}
	\caption{Schematic diagram presented by \cite{ramsay1982} to explain ``the concept of a functional datum.'' The diagram appeared as Figure~2 of that paper above the caption: ``Possible domains for statistical observations. These domains depend on whether the number of replications or
		cases ($n$) is finite or infinite and whether the number of variables or points of observation ($p$) is finite or corresponds to the points on a continuum.'' The upper right represents functional data, whereas the lower left scenario, mentioned but not developed by \cite{ramsay1982}, is the continuous-time multivariate analysis setup of the present paper. Figure reproduced with permission from Springer Nature.}\label{spring}
\end{figure}

Heuristically, our proposed framework, henceforth \emph{continuous-time multivariate analysis} (CTMVA), is the ``transpose'' of functional data analysis \citep[FDA;][]{ramsay2005}, in the following sense. In a programmatic address that introduced the term ``functional data'' and foreshadowed the FDA paradigm, 
\cite{ramsay1982} argued that some modern data sets call for replacing the traditional $p$ variables with infinitely many points on a continuum, as in the upper right picture in \figref{spring}, and suggested that spline methodology could play a key role in handling such infinite-dimensional data. He briefly noted that instead of taking $p=\infty$ as in FDA, one might consider situations in which $n=\infty$, as in the lower left picture in \figref{spring}; but he did not develop the $n=\infty$ case. The CTMVA framework is an attempt to do so, while maintaining an emphasis on the role of spline and other basis-function representations as in \cite{ramsay1982} and \cite{ramsay2005}.

\begin{figure}
	\centering 
	\includegraphics[width=.8\textwidth]{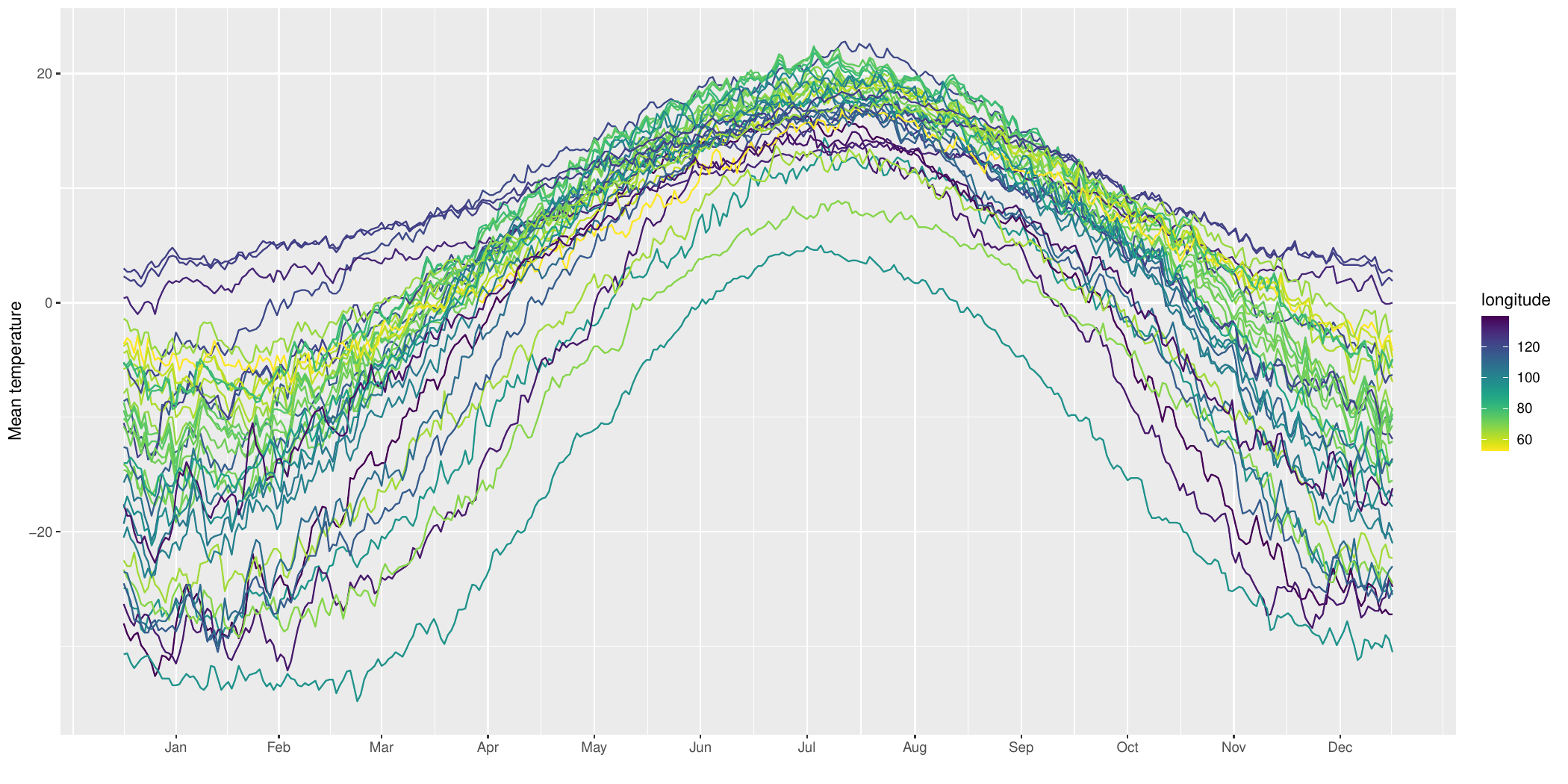}
	\includegraphics[width=.8\textwidth]{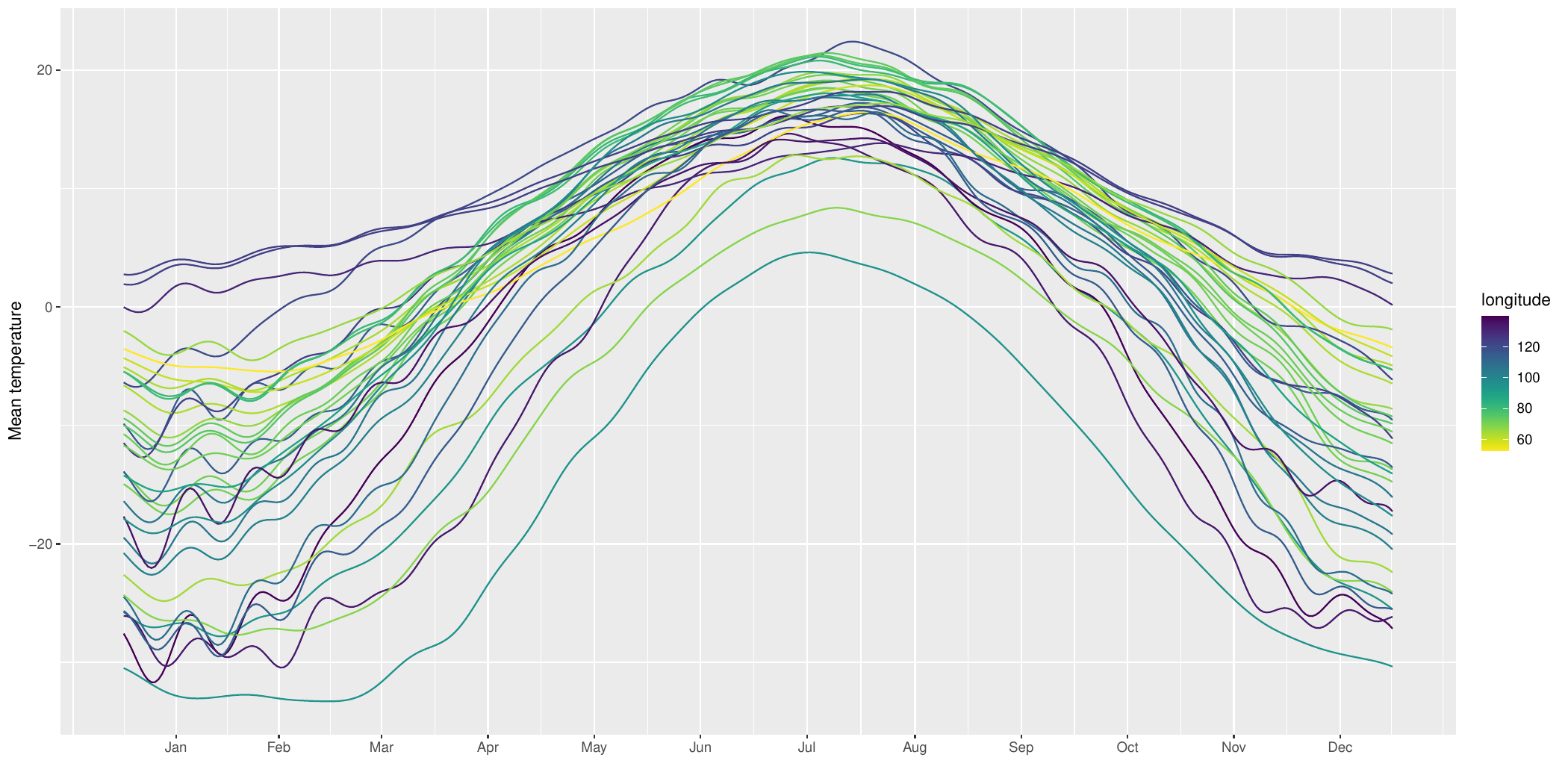}
	\includegraphics[width=.89\textwidth]{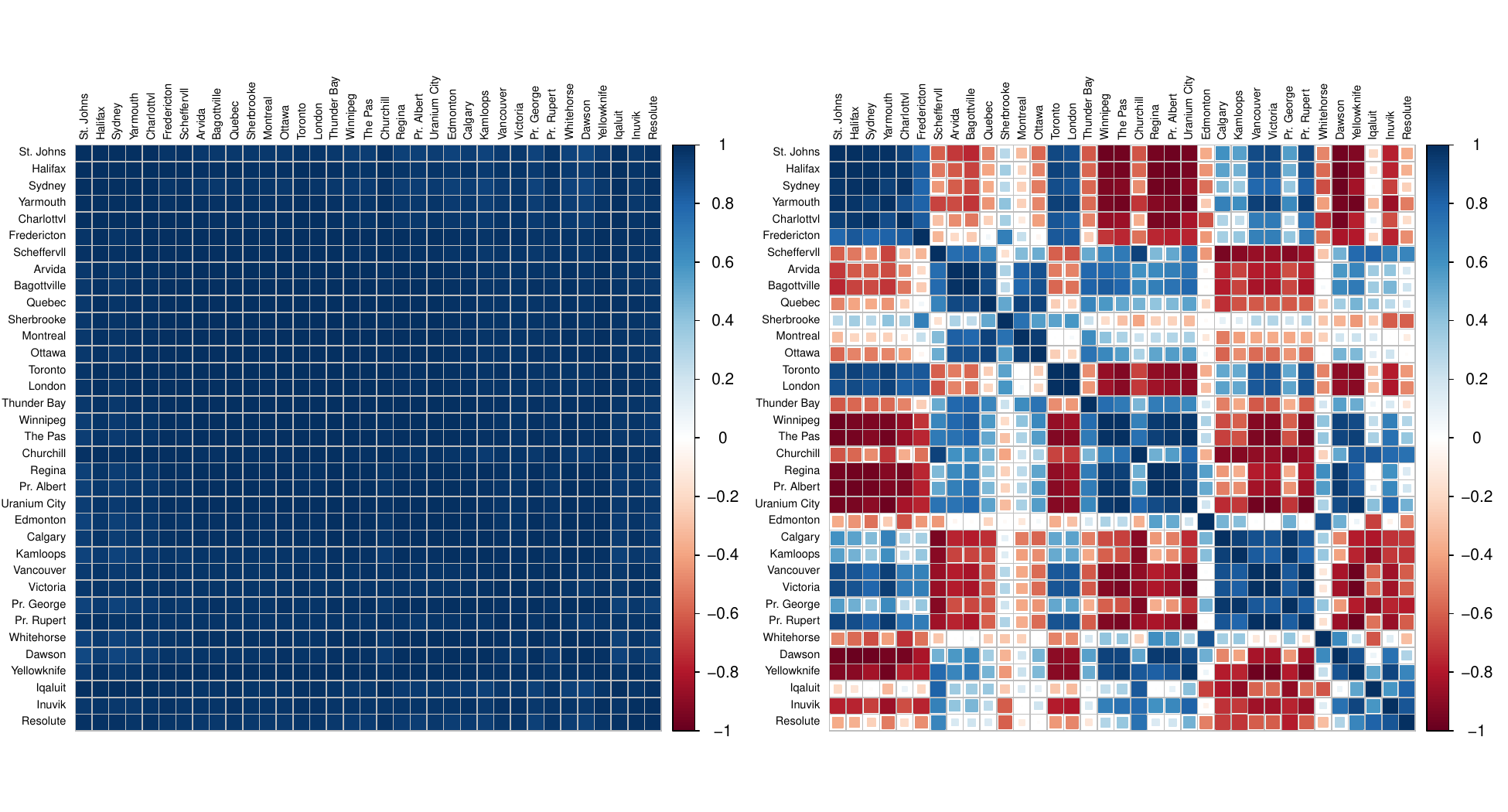}
	\caption{Top: Mean daily temperatures at 35 Canadian weather stations \citep{ramsay2005}. Middle: The same data after smoothing with respect to a Fourier basis.  Below: Continuous-time correlations among daily mean temperatures in the 35 weather stations, visualized by the corrgram method \citep{friendly2002} implemented in \cite{corrplot}, before (left) and after (right) removing the common seasonal trend.}\label{tempcor}
\end{figure}

To illustrate the contrast between CTMVA and FDA it may be instructive to reconsider what is perhaps the most famous example of functional data: the Canadian temperature data of \cite{ramsay2005}, consisting of mean tempera\-tures on each day of the year at 35 weather stations in Canada. In FDA, these are viewed not as $n=35$ observations of dimension $p=365$, but as smooth curves $x_1(t),\ldots,x_{35}(t)$ on the time interval $\cI=[0,365]$, so that in effect $p=\infty$ (uncountably infinite). \figref{tempcor} displays both the raw data (upper panel) and the curves following roughness-penalty smoothing with respect to a 45-element Fourier basis (middle panel), with the smoothing parameter chosen by restricted maximum likelihood \citep{ruppert2003,reiss2009,wood2011} .
In CTMVA we again view the data as smooth curves $x_1(t),\ldots,x_{35}(t)$, but these are now understood as $n=\infty$ observations of $p=35$ variables.  

In FDA, given a sample of curves arising from a stochastic process on a finite interval $\cI\subset\mathbb{R}$, much interest focuses on $\Gamma(s,t)$, the covariance function of the underlying process. In particular, in functional principal component analysis one estimates the eigenfunctions of the associated covariance operator. In CTMVA, by contrast, the covariances or correlations of interest are those among the curves---representing, in this example, relation\-ships among the various weather stations.  In \secref{2forms} below we discuss the matrices shown at the bottom of \figref{tempcor}, representing correlations among the 35 stations' temperature curves, and in Appendix \ref{dapic} we consider correlations among precipitation curves from the same data set.

The practical impetus for CTMVA is twofold. First, the strategy we present below, of representing a multivariate time series as a set of smooth curves rather than a discrete set of MV observations, can sometimes improve performance of MVA techniques. We demonstrate this below for correlation estimation, in the simulation study of \secref{simsec}, and for clustering, with the environmetric data analyzed in \secref{chisec}. Second, CTMVA can sometimes be applied when classical MVA cannot---for example, if the $p$ variables are not measured at a uniform set of time points (as in \secref{wdisec} below).

After presenting the basic setup and notation in \secref{setup}, in \secref{covmat} we define continuous-time versions of the sample mean vector and the sample covariance and correlation matrices. As part of this development, \secref{2forms} clarifies the basic but possibly confusing role of centering, which can have an important effect on correlation estimation. In \secref{pca}--\secref{kmeans} we develop continuous-time (CT) versions of three key MVA techniques, namely principal component analysis, Fisher's linear discriminant analysis and $k$-means clustering. A simulation study and an application to air pollution data are presented in \secref{simsec} and \secref{chisec}, respectively. The paper concludes with a brief discussion in \secref{discuss}. 

\section{Setup}\label{setup}
In what follows we assume $\bx(t)\equiv[x_1(t),\ldots,x_p(t)]^T$ to be a realization of a stochastic vector process $\cX:\cI\rightarrow\mathbb{R}^p$. Depending on the application, the process $\cX(t)=[X_1(t),\ldots,X_p(t)]^T$ may or may not be stationary. If it is, the mean vector 
$\bmu(t) = E[\cX(t)]\in\mathbb{R}^p$ and covariance matrix $\bSigma(t)=E[\{\cX(t)-\bmu(t)\}\{\cX(t)-\bmu(t)\}^T]\in\mathbb{R}^{p\times p}$ do not depend on $t$ and thus may be denoted by $\bmu,\bSigma$.  In some applications the $p$ functions exhibit a common trend, i.e.\ a  time-varying smooth function $m(t)=E[\frac{1}{p}\sum_{u=1}^p X_u(t)]$, an estimate of which should be subtracted from each component of $\bx(t)$ before proceeding with data analysis. We explain this with reference to an example in \secref{2forms}.

Following \cite{ramsay2005}, we make the fundamental simplifying assumption that, for $u=1,\ldots,p$,  the $u$th function can be represented by means of an expansion $x_u(t)=\bc_u^T\bphi(t)$ where 
$\bphi(t)=[\phi_1(t),\ldots,\phi_K(t)]^T$   is a vector of \emph{a priori} smooth basis functions on $\cI$.  Thus, if  $\bC$ is the $K\times p$ coefficient matrix with $u$th column $\bc_u$, then 
\begin{equation}\label{fbo}\bx(t)=\bC^T\bphi(t).\end{equation}
Advantages of basis functions over other smoothing techniques include flexibility, compact representation of even large MV data sets, and not requiring equally spaced time points or a uniform set of times for the component functions.

The observed data from which this basis expansion is derived may be a traditional $n\times p$ data matrix, representing $p$ variables observed at a common set of $n$ time points, with smoothing applied to each of the $p$ columns (see \secref{estest} below). But in some applications (e.g., in \secref{wdisec}), the variables may be observed at different time points. The only requirement here is that the same basis is employed for all $p$ curves. For  curves that are cyclic it is natural for $\bphi$ to be a Fourier basis: for  the weather data, for example, this ensures fits that transition smoothly from December 31 to January 1. For other applications, $B$-spline bases, most often cubic $B$-splines, are a very popular choice due to their computational efficiency and favorable approximation properties. In practice, the representation is typically obtained by penalized smoothing (see \secref{penalize}), and this renders the results fairly insensitive to the choice of $K$ \citep[e.g.,][]{ruppert2003}.

Our implementation of CTMVA in the R environment \citep{R} depends on the $B$-spline and Fourier basis functions from the \texttt{fda} package for functional data analysis \citep{ramsay2009,fda}. In some applications it might be desirable to employ data-driven basis functions such as functional principal components \citep{silverman1996}.

If $\cS$ is a subinterval, or finite union of subintervals, of $\cI$, we define 
\begin{equation}\label{bmv}
	\bar{\bphi}_\cS=|\cS|^{-1}\int_\cS\bphi(t)dt\in\mathbb{R}^K\quad\mbox{and}
	\quad\bQ_\cS=|\cS|^{-1}\int_{\cS}[\bphi(t)-\bar{\bphi}_\cS][\bphi(t)-\bar{\bphi}_\cS]^Tdt\in\mathbb{R}^{K\times K}.
\end{equation}
Let $\bar{\bphi}=\bar{\bphi}_\cI$ and $\bQ=\bQ_\cI$. These expressions will be used repeatedly in what follows. Computation of $\bar{\bphi}$ and $\bQ$ is discussed in Appendix~\ref{compdet}.

\section{Continuous-time sample mean vector and covariance and correlation matrices}\label{covmat}

\subsection{Definitions}\label{defsec}
In what follows we motivate the continuous-time sample mean vector and sample covariance matrix as analogues, or limits, of their classical counterparts. These will be seen, moreover, to be natural  estimates of corresponding stochastic process parameters  (see \secref{estest} below).

In classical multivariate analysis, given a data matrix 
$\bX=(x_{ij})_{1\leq i\leq n, 1\leq j\leq p}$
representing $n$ observations of a $p$-dimensional random vector, the $p$-dimensional sample mean vector and $p\times p$ covariance matrix are given, respectively, by $\bar{\bx}=(\bar{x}_1,\ldots,\bar{x}_p)^T$ with $\bar{x}_u=n^{-1}\sum_{i=1}^nx_{iu}$ and
$\bS=\left[\frac{1}{n}\sum_{i=1}^n(x_{iu}-\bar{x}_u)(x_{iv}-\bar{x}_v)\right]
_{1\leq u,v\leq p}$  [here  we follow \cite{mardia1979} in dividing by $n$ rather than by $n-1$].  

The CT (sample) mean vector $\bar{\bx}^*$ and covariance matrix $\bS^*$ (here and below we use $^*$ to signify CT analogues of classical MVA estimators)    are simply the limits, as $n\rightarrow\infty$, of their classical counterparts $\bar{\bx},\bS$, when the rows of $\bX$ are $\bx(t_1)^T,\ldots,\bx(t_n)^T$, where $t_1,\ldots,t_n$ form a grid of points spanning $\cI$ at equal intervals of $|\cI|/n$ (e.g., $t_i=i/n$ for each $i$, for $\cI=[0,1]$). In that case $|\cI|\bar{\bx}=\frac{|\cI|}{n}\sum_{i=1}^n\bx(t_i)$ is  a Riemann sum converging to $\int_\cI\bx(t)dt$ as $n\rightarrow\infty$; dividing by $|\cI|$ yields $\bar{\bx}\rightarrow\bar{\bx}^*$, where \begin{equation}\label{meandef}\bar{\bx}^*=(\bar{x}^*_1,\ldots,\bar{x}^*_p)^T\quad\mbox{with}\quad\bar{x}^*_u=|\cI|^{-1}\int_{\cI}x_u(t)dt.\end{equation}
Consequently, 
$|\cI|\bS=\frac{|\cI|}{n}\sum_{i=1}^n[\bx(t_i)-\bar{\bx}][\bx(t_i)-\bar{\bx}]^T$ is asymptotically equivalent to $\frac{|\cI|}{n}\sum_{i=1}^n[\bx(t_i)-\bar{\bx}^*][\bx(t_i)-\bar{\bx}^*]^T$,
which  is a Riemann sum converging to $\int_{\cI}[\bx(t)-\bar{\bx}^*][\bx(t)-\bar{\bx}^*]^Tdt$. Dividing by $|\cI|$ leads to $\bS\rightarrow\bS^*$, where
\begin{equation}\label{cts}\bS^*=\left[|\cI|^{-1}\int_{\cI}[x_u(t)-\bar{x}^*_u]][x_v(t)-\bar{x}^*_v]dt\right]_{1\leq u,v\leq p}=|\cI|^{-1}\int_{\cI}[\bx(t)-\bar{\bx}^*][\bx(t)-\bar{\bx}^*]^Tdt.\end{equation}
The basis representation \eqref{fbo} makes the integrals in \eqref{meandef}, \eqref{cts} computable even though $\bx(t)$ is observed at only finitely many times $t$,  and \eqref{bmv} yields convenient and succinct expressions for the CT mean and covariance matrix: $\bar{\bx}^*=\bC^T\bar{\bphi}$ and $\bS^*=\bC^T\bQ\bC$. 

The CT (sample) correlation between the $u$th and $v$th functions is similarly defined as 
\begin{equation}r^*_{uv}=\frac{\int_{\cI}x^c_u(t)x^c_v(t)dt} {\left[\int_{\cI}x^c_u(t)^2dt\int_{\cI}x^c_v(t)^2dt\right]^{1/2}},\label{cordef}\end{equation}
where $x^c_u(t)=x_u(t)-\bar{x}^*_u$. As in classical MVA, the matrix of CT correlations equals 
$\bR^*=\bD^{*-1/2}\bS^*\bD^{*-1/2}$,
where $\bD^*$ is the diagonal matrix with the same main diagonal as $\bS^*$.

\subsection{An example: World Development Indicators}\label{wdisec}
To illustrate the use of CT correlation for time series with irregularly spaced times and substantial missingness, 
we consider several of the World Development Indicators (WDI; \url{https://datacatalog.worldbank.org}), variables that are estimated annually for each nation in the world. 
Quantitative political scientists use such national time series to investigate relationships among different indicators.  Missing values are common, and may be addressed with ad hoc fixes \citep[e.g.,][]{baum2003} or complex imputation schemes \citep[e.g.,][]{honaker2010}. CT correlation offers a simple and fast way to assess relationships between a potentially large number of indicators, while allowing for substantial missing data. Such applications of CT correlation may prove highly useful for exploratory data analysis and hypothesis generation, which are increasingly important goals in light of the popularity of online resources such as Gapminder and Our World in Data, and the large number of available candidate variables.

\begin{figure}
	\centering
	\includegraphics[width=.49\textwidth]{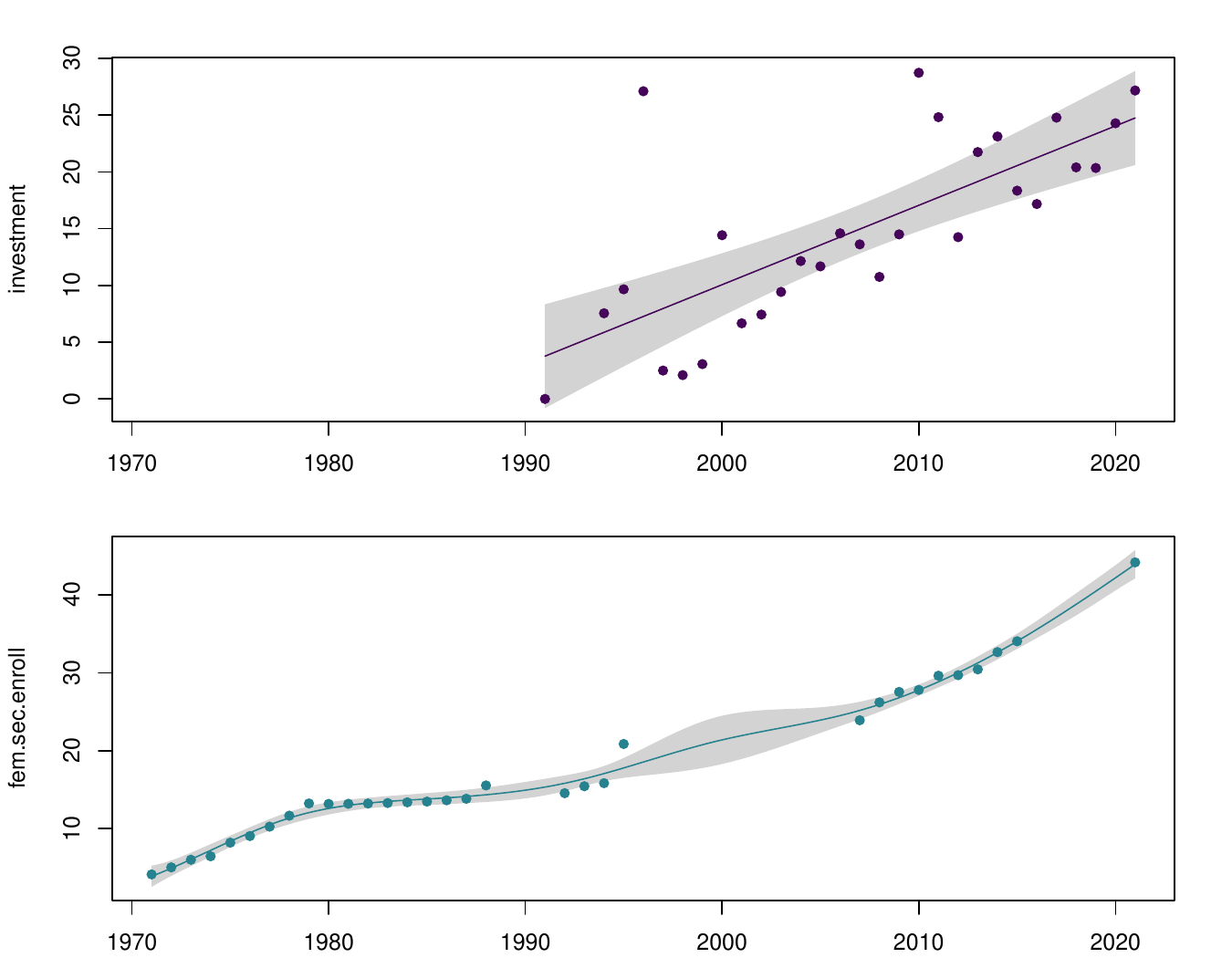}
	\includegraphics[width=.49\textwidth]{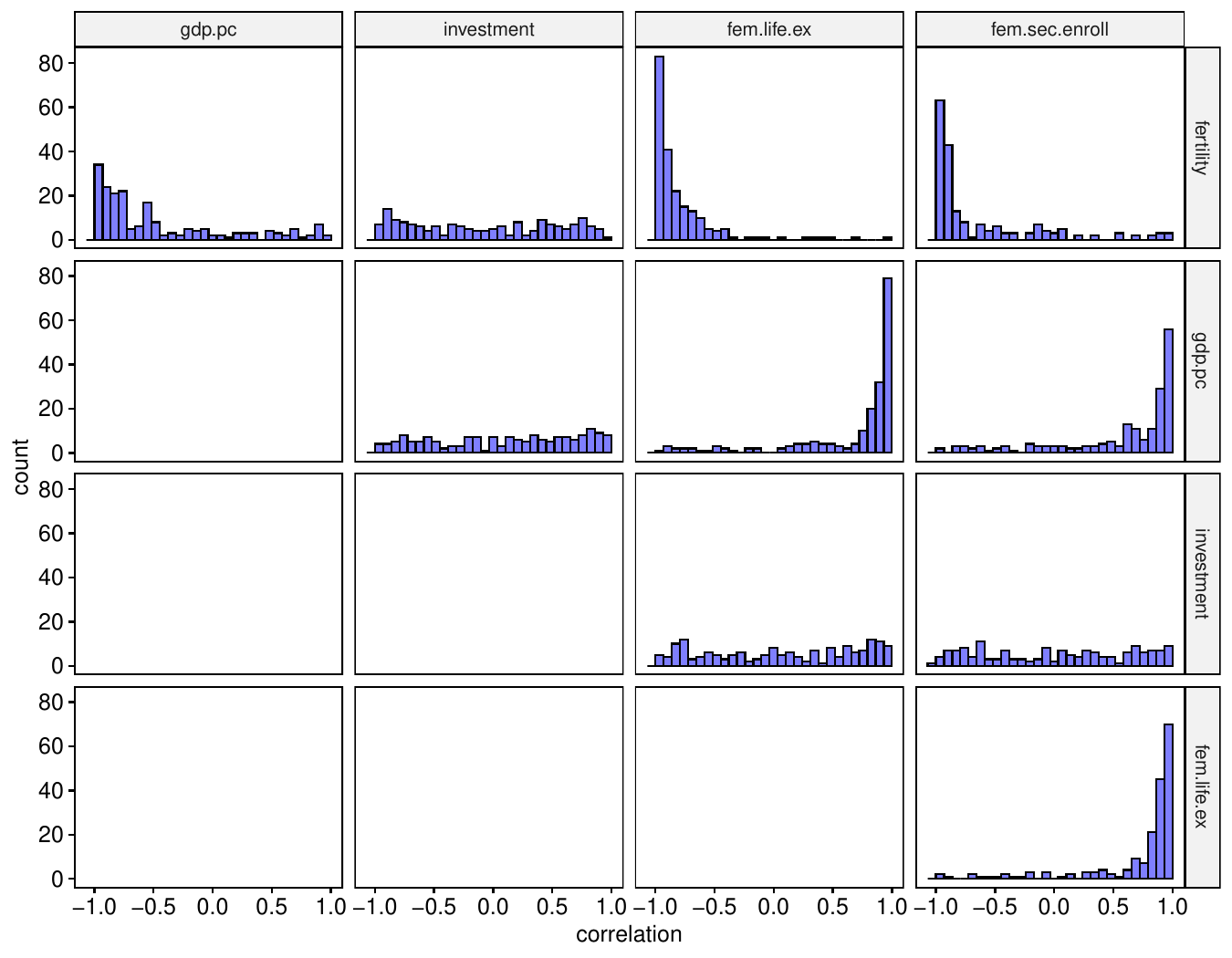}
	\caption{Left: Domestic fixed investment (above) and female secondary school enrollment (below) for DR Congo, with spline smooths $\pm$ 2 standard errors. CT correlation is computed for 1991--2021, the overlap of the two variables' time ranges, resulting in an estimate of 0.977. Right: Histograms of country-level CT correlations for each pair of variables among the five WDIs considered.}\label{agofig}
\end{figure}

We focus here on five variables taken from \cite{baum2003}: fertility, gross domestic product (GDP) per capita, gross domestic fixed investment, female life expectancy, and female secondary school enrollment. At left in \figref{agofig} are plots of investment and female enrollment in the Democratic Republic of the Congo, which illustrate the difficulty of assessing correlation between highly incomplete time series. In only 12 years are both variables estimated, so that ordinary correlation can be computed with only 12 observations. But the CT correlation is obtained by using all observations for 1991--2021, the overlap of the two variables' time ranges, to estimate the trend in each variable.  %

The resulting correlation estimate of 0.977 is not very precise, with a 95\% confidence interval from 0.500 to 0.997 based on 999 bootstrap replicates. (In \secref{discuss} we mention a weighting approach that might help to improve the precision of CT correlation.)
But while individual correlation estimates from this data set lack precision, it is possible to discern interesting patterns from a collection of such estimates.
We computed CT correlation for each pair of variables, for each of the 216 countries in the data set, subject to the requirements that (i) each variable have at least 8 observations and (ii) the overlap of the two time ranges be at least 12 years. These requirements were met for 85\% of the $216 {5 \choose 2}$ possible correlations, resulting in 1840 CT correlations, which were computed in 27 seconds on a MacBook Pro with 32 GB of RAM. At right in \figref{agofig} are histograms of the country-specific CT correlations for each pair of variables. Notwithstanding the above results for DR Congo, there is no clear international pattern of positive or negative correlation between investment and any of the other variables. On the other hand, the four histograms in the top right corner suggest that in most countries, historical trends in two measures of women's development, female life expectancy and female school enrollment, are negatively correlated with fertility trends and positively correlated with GDP trends.

\subsection{Estimators or estimands?}\label{estest}
In classical  MVA, the sample mean $\bar{\bx}$ is an estimate, based on 
observations $\bx_1,\ldots,\bx_n$ from a common multivariate distribution, of the population mean $\bmu=E(\cX)$ of a random vector $\cX$ arising from that distribution, while the sample covariance matrix  $\bS$ is an estimate of the population covariance matrix $\bSigma=E[(\cX-\bmu)(\cX-\bmu)^T]$. 
Likewise, in CTMVA, $\bar{\bx}^*,\bS^*$ are natural estimators of  $\bmu,\bSigma$ as defined in \secref{setup}, assuming stationarity. If the process is not stationary, $\bar{\bx}^*,\bS^*$ can be viewed as estimators of $|\cI|^{-1}\int_\cI\bmu(t)dt$ and $|\cI|^{-1}\int_\cI\bSigma(t)dt$, respectively.

But in many applications, interest centers not on the underlying process but on the specific instance $\bx(\cdot)$. It is then more appropriate to regard $\bar{\bx}^*,\bS^*$, as well as $\bR^*$, not as \emph{estimators} but as \emph{estimands}, since in practice the function $\bx(\cdot)$ on which they are based is not observed perfectly, for three reasons: (i) the $p$ component functions are observed only at finitely many time points, (ii) they are generally observed with noise and/or measurement error, and (iii) the basis representation \eqref{fbo} introduces approximation error.

Formally, for $u=1,\ldots,p$, at finitely many times $t_{iu}$ we observe not the true $x_u(t_{iu})$ but 
\begin{equation}\label{zij}z_{iu}=x_u(t_{iu})+\varepsilon_{iu},\end{equation} where $\varepsilon_{iu}$ represents zero-mean error. We apply penalized smoothing \citep{ruppert2003,wood2017} to the $z_{iu}$'s for each $u$, to obtain estimates  $\hat{\bx}(t)=\widehat{\bC}^T\bphi(t)\in\mathbb{R}^p$ for the $p$ functions. We note that for given $u$, the  $\varepsilon_{iu}$'s may not be independent. The penalized spline literature includes methods to take residual dependence into account, as well as some evidence that likelihood-based smoothness selection is robust to such dependence \citep{krivobokova2007}. 
In this setup, $\bar{\bx}^*, \bS^*,\bR^*$ are unobservable but are estimated, respectively, by 
\[\hat{\bar{\bx}}^*=|\cI|^{-1}\int_{\cI}\hat{\bx}(t)dt=\widehat{\bC}^T\bar{\bphi},\quad\quad
\widehat{\bS}^*=|\cI|^{-1}\int_{\cI}[\hat{\bx}(t)-\hat{\bar{\bx}}^*][\hat{\bx}(t)-\hat{\bar{\bx}}^*]^Tdt=\widehat{\bC}^T\bQ\widehat{\bC},\]
and $\widehat{\bR}^*=\widehat{\bD}^{*-1/2}\widehat{\bS}^*\widehat{\bD}^{*-1/2}$,  where $\widehat{\bD}^*$ is the diagonal matrix with the same main diagonal as $\widehat{\bS}^*$. For simplicity, in what follows we omit the ``hats'' and treat $\bx$ and consequently $\bar{\bx}^*,\bS^*,\bR^*$ as given, except when presenting the simulation study of \secref{simsec}, which shows the advantage of CT over ordinary correlation for stochastic processes observed with noise.\label{penalize}

To restate the above argument more succinctly and concretely, let us consider a bivariate process that is stationary, so that the covariance and hence the correlation $\rho$ between the two components are time-independent.   Classical MVA distinguishes the population parameter $\rho$ from the sample quantity $r$. But in CTMVA there are not two but three quantities: the process correlation $\rho$ gives rise to the CT correlation $r^*$ between a pair of functions drawn from the process, which is estimated from data by $\hat{r}^*$---or schematically, \begin{equation}\label{3step}\rho\xrightarrow{(i)} r^* \xrightarrow{(ii)} \hat{r}^*\end{equation}
(where the arrows informally denote ``gives rise to'' rather than denoting convergence).
Depending on the context we may be more interested in step (i) or in step (ii) of \eqref{3step}.

\section{Two forms of centering}\label{2forms}

The mean-centered variable vector $\bx^c(t)=\bx(t)-\bar{\bx}^*$ appeared in \secref{defsec} and has a number of other uses, such as for CT principal component analysis (see \secref{pca} below). By \eqref{fbo}, $\bx^c(t)=\bC^T[\bphi(t)-\bar{\bphi}]$, i.e.\ the centered variables are linear combinations of $\phi_1(t)-\bar{\phi}_1,\ldots,\phi_K(t)-\bar{\phi}_K$, but it would be more convenient if we could represent them as linear combinations of the original (not centered) basis functions $\phi_1(t),\ldots,\phi_K(t)$. It turns out that we can do so, under the simple condition that \begin{equation}\label{wphi}\mbox{there exists }\bw\in\mathbb{R}^K\mbox{ such that }\bw^T\bphi(t)=1\mbox{ for all }t,\end{equation}
since then $\bx^c(t) =\bC^T[\bphi(t)-\bar{\bphi}] =\bC^T[\bI_K-\bar{\bphi}\bw^T]\bphi(t)$, 
i.e., the centered curves have a basis representation as in  \eqref{fbo}, but with coefficient matrix $[\bI_K-\bw\bar{\bphi}^T]\bC$ in place of $\bC$. Condition \eqref{wphi} holds for $B$-spline bases, with $\bw=(1,\ldots,1)^T$, and for Fourier bases, with $\bw=(|\cI|^{1/2},0,\ldots,0)^T$.

Centering in the above sense is the continuous-time version of column-centering an $n\times p$ data matrix. As alluded to in \secref{setup}, in some applications, a CT analogue of \emph{row}-centering is warranted, i.e., an estimated common trend\label{common} $\widehat{m}(t)=\frac{1}{p}\sum_{u=1}^p x_u(t)$ should be subtracted from each of the $p$ curves before computing the co\-vari\-ance or correlation. Consider, for example, the Canadian temperature curves in \figref{tempcor}.   Without removal of the common trend,   the CT correlation matrix $\bR^*$ for this example has all elements above 0.9, since the temperatures at all stations are relatively high in the summer and low in the winter, yielding  the uninformative correlation matrix in the lower left panel of \figref{tempcor}.    But with detrending applied, the result is the more interesting correlation matrix in the lower right panel.  The 35 stations are listed geographically, roughly from east to west, except for the last six stations, which are located in the Canadian North. The top and middle panels of \figref{tempcor} suggest that in many cases, stations with nearly the same longitude have similar temperature curves. This explains why the CT correlation matrix has several blocks of very high positive correlations along the main diagonal: for instance, the first six stations are those located in the Atlantic provinces, and the correlations among  them are all above 0.78 and mostly above 0.93.
In Appendix \ref{diagnose} we propose a simple diagnostic to help determine, for a given data set, whether detrending is advisable.

Since the Canadian weather curves are spatially indexed, \cite{scheipl2015} modeled the spatial correlation to improve precision in an FDA context \citep[see also][]{liu2017}. In the above analysis, however, the correlations among stations are the estimands of interest, rather than a means to a further inferential end as in spatial FDA.

\section{CT principal component analysis}\label{pca}

In classical MVA, the (sample) principal components are linear combinations $\bv_1^T\bx,\ldots,\bv_p^T\bx$ that have maximal variance, subject to mutual orthonormality of the $\bv_j$'s.
If the sample covariance matrix
$\bS$ has orthonormal eigenvectors $\be_1,\ldots,\be_p$ with corresponding eigenvalues $\lambda_1\geq\ldots\geq\lambda_p$, then by a standard result on Rayleigh quotient maximization, the solution to this iterative optimization problem is $\bv_m=\be_m$ for $m=1,\ldots,p$. 

Continuous-time PCA seeks linear combination \emph{functions} $\bv_m^T\bx(t)$ that achieve maximal variance in the above  sense, but with the variance of $\bv^T\bx(t)$ defined as in \eqref{cts}  with respect to time, as  $|\cI|^{-1}\int_\cI [\bv^T\bx^c(t)]^2dt = \bv^T\bS^*\bv$.   By the same argument as in the classical case, but with $\bS^*$ in place of $\bS$, the solution is given by the eigenvectors of $\bS^*$: that is, $\bv_m=\be^*_m$ for $m=1,\ldots,p$, where $\be^*_1,\ldots,\be^*_p$ are orthonormal eigenvectors of $\bS^*$ in descending order of the corresponding eigenvalues $\lambda^*_1\geq\ldots\geq\lambda^*_p$. 

Similarly to the classical setting, we then have the expansion
$\bx^c(t)=\sum_{j=1}^p \be^*_js_j(t)$,
where $s_j(t)$ is the $j$th score function, defined by $s_j(t)=\be^{*T}_j\bx^c(t)$,
and it follows straightforwardly from this definition that the CT covariance matrix \eqref{cts} of $[s_1(t),\ldots,s_p(t)]^T$ is $\mbox{Diag}(\lambda^*_1,\ldots,\lambda^*_p)$. Assuming \eqref{wphi}, since each component of $\bx^c(t)$ is a linear combination of the basis functions $\phi_1(t),\ldots,\phi_K(t)$, the same is true for each of the scores $s_1(t),\ldots,s_p(t)$. Appendix \ref{eegsec} presents an application of CT PCA to electroencephalography data.

Canonical correlation analysis (CCA) is another classical MVA method that is somewhat related to PCA. Here one has two sets of variables and seeks pairs of linear combinations, one for each set, having maximal correlation in an iterative sense. A continuous-time version of CCA is presented in Appendix~\ref{cca}.

\section{CT Fisher's linear discriminant analysis}\label{lda}

In its classical form, Fisher's linear discriminant analysis (LDA; see Appendix~\ref{flda}) seeks linear combinations $\bv^T\bx$ that optimally separate $G\geq 2$ sets of observations $\bX_1,\ldots,\bX_G$ into which the data set is divided \emph{a priori}. 
Like PCA and CCA, Fisher's LDA finds the desired linear combinations of the variables by solving an iterative optimization problem, in this case based on a partition of the total sums of squares and cross-products matrix $\bT=n\bS$ into \emph{within-groups} and \emph{between-groups} components, $\bT=\bW+\bB$. 

We propose a continuous-time variant of Fisher's LDA that optimally discriminates among $G\geq 2$ \emph{subintervals} $\cI_1,\ldots,\cI_G$ into which $\cI$ is partitioned. For $g\in\{1,\ldots,G\}$, let  $\bar{\bx}^*_g=\bC^T\bar{\bphi}_{\cI_g}=|\cI_g|^{-1}\int_{\cI_g}\bx(t)dt$. Analogously to the classical setting, we have the decomposition   $\bT^*=\bW^*+\bB^*$,  where
\begin{eqnarray*}
	\bT^*&=&\int_{\cI}[\bx(t)-\bar{\bx}^*] [\bx(t)-\bar{\bx}^*]^Tdt,\\
	\bW^*&=&\sum_{g=1}^G\int_{\cI_g}[\bx(t)-\bar{\bx}^*_g] [\bx(t)-\bar{\bx}^*_g]^Tdt,\\
	\bB^*&=&\sum_{g=1}^G|\cI_g|(\bar{\bx}^*_g-\bar{\bx}^*)(\bar{\bx}^*_g-\bar{\bx}^*)^T.
\end{eqnarray*}
(see Appendix \ref{flda} for a derivation). The CT Fisher's linear discriminants are then  functions $\bv_1^T\bx(t),\ldots,\bv_s^T\bx(t)$ that optimally separate the subintervals in the sense of iteratively maximizing $\bv^T\bB^*\bv / \bv^T\bW^*\bv$, subject to $\bv_i^T\bW^*\bv_j=0$ for $i\neq j$. By the same argument as in the classical setting, the optimally separating vectors $\bv_1,\ldots,\bv_s$  are eigenvectors of $\bW^{*-1}\bB^*$ corresponding to the positive eigenvalues in descending order. 
Once again, our implementation relies on the basis representation \eqref{fbo} along with \eqref{bmv}:  we have
\begin{eqnarray}
	\bT^*&=&|\cI|\mbox{ }\bC^T\bQ\bC,\nonumber\\
	\bW^*&=&\bC^T\left(\sum_{g=1}^G|\cI_g|\bQ_{\cI_g}\right)\bC, \label{wstar}\\
	\bB^*&=&\bC^T\left[\sum_{g=1}^G|\cI_g|(\bar{\bphi}_{\cI_g}-\bar{\bphi})(\bar{\bphi}_{\cI_g}-\bar{\bphi})^T\right]\bC.\nonumber
\end{eqnarray}

\section{CT $k$-means clustering}\label{kmeans}
Continuous-time  $k$-means clustering partitions $\cI$ into \emph{temporal clusters}, each of which is  a subinterval, or a finite union of subintervals, of $\cI$. Our algorithm is a CT extension of the basic $k$-means procedure introduced in a 1957 manuscript later published as \cite{lloyd1982}:
\begin{enumerate}
	\item Randomly select centers $\bm_1,\ldots,\bm_k\in\mathbb{R}^p$, by evaluating $\bx(t)$ at $k$ randomly chosen time points.
	\item Until a convergence criterion is met:
	\begin{enumerate}
		\item Partition $\cI$ into clusters $\cC_1,\ldots,\cC_k$ defined by $\cC_i=\{t\in\cI:d_i(t)<d_j(t)\mbox{ for all }j\neq i\}$
		for each $i$, where $d_i(t)=\|\bx(t)-\bm_i\|$.
		\item For $i=1,\ldots,k$, update $\bm_i=|\cC_i|^{-1}\int_{\cC_i}\bx(t)dt=\bC^T\bar{\bphi}_{\cC_i}$.
	\end{enumerate}
\end{enumerate}

The nontrivial part is step 2(a). Since $d_i(t)^2=\|\bC^T\bphi(t)-\bm_i\|^2=\bphi(t)^T\bC\bC^T\bphi(t)-2\bphi(t)^T\bC\bm_i+\bm_i^T\bm_i$, whose first term does not depend on $i$, it follows that finding the cluster center nearest to $\bx(t)$ reduces to finding the $i\in\{1,\ldots,k\}$ that minimizes
\[A_i(t)\equiv -2\bphi(t)^T\bC\bm_i+\bm_i^T\bm_i.\] 
Suppose for simplicity that $k=2$. Then each of $\cC_1=\{t\in\cI:A_1(t)<A_2(t)\}$ and $\cC_2=\{t\in\cI:A_1(t)>A_2(t)\}$  is a union of subintervals of $\cI$, where the boundaries between these subintervals are the points $t$ at which $A_1(t)$ and $A_2(t)$ cross, i.e., the zero crossings of 
\begin{equation}\label{a1a2}A_1(t)-A_2(t)=2\bphi(t)^T\bC(\bm_2-\bm_1)+\bm_1^T\bm_1-\bm_2^T\bm_2.\end{equation}
If the basis functions are cubic splines then this expression   is a cubic polynomial in $t$ on each of the intervals between knots. Thus finding the transition points between the two clusters reduces to finding, for the cubic polynomial on each of these intervals, any zero crossings that occur within that interval.  In this way the problem of partitioning infinitely many time points among $k$ clusters is reduced to identifying the zeros of a finite set of polynomials. Appendix~\ref{kfast} provides a formula for the polynomial coefficients, and discusses how this CT approach to $k$-means clustering may improve computational efficiency.  The $k>2$ case is similar to the above, but not every crossing point of any pair of $A_i(t)$'s is then a transition point. 

Analogously to ordinary $k$-means, a CT $k$-means clustering decomposes the total integrated sum of squared distances from $\bar{\bphi}$ into between-cluster and within-cluster components. The latter component, more precisely the total within-cluster integrated squared distance from the cluster mean, is the objective function of CT $k$-means and is readily shown to equal
$\tr\left[\bC^T\left(\sum_{i=1}^k|\cC_i|\bQ_{\cC_i}\right)\bC\right]$  [cf.\ \eqref{wstar}].  Like ordinary $k$-means, the above CT algorithm is not guaranteed to converge to the global minimum of this objective function. It is thus recommended to run the algorithm repeatedly with different starting values, and choose the solution that minimizes the objective.

In Appendix \ref{sil} we present a continuous-time variant of the silhouette width \citep{rousseeuw1987}, which indicates how clearly a given time point belongs to its assigned cluster. The silhouette width can be plotted for each of a grid of time points, and its average value provides a simple index of how tight the clusters are. As with ordinary $k$-means clustering, the average silhouette width may be used to guide the choice of $k$. 

\section{A simulation study}\label{simsec}
We conducted a simulation study to assess the performance of CT correlation estimation relative to classical (discrete-observation) correlation between $p=2$ variables.  True curves $\bx(t)$ were generated from a bivariate Gaussian process on $[0,1]$ (see Appendix \ref{simdet}) with between-curve covariance matrix $\bSigma=\left(\begin{array}{cc} 1 & 0.5 \\ 0.5 & 1\end{array}\right)$,   and thus correlation $\rho=0.5$, and within-curve covariance function $\Gamma$ having the squared exponential (radial basis function) form 
\begin{equation}\label{squexp}
	\Gamma(s,t)=\exp[-(s-t)^2/(2\ell^2)]\end{equation}
where the \emph{length parameter} $\ell>0$ determines the wiggliness of the curves, with higher $\ell$ implying greater smoothness.   
For each pair of curves we obtain noisy observations   \eqref{zij} at each of 500 common observation times, with noise standard deviation $\sigma=0.5$.
For a given pair of curves $\bx(\cdot)$, the true CT correlation between them is $r^*=r^*_{12}$ [see \eqref{cordef}];    the CT correlation \emph{estimate}, based on an estimate $\hat{\bx}(\cdot)$ of the pair of curves, is denoted by $\hat{r}^*$ (see \secref{estest}).   We compare the ordinary correlation $\hat{r}$, based on $n=500$ bivariate observations, versus CT correlation $\hat{r}^*$, based on treating the number of observations $n$ as infinite and smoothing both curves with respect to a $B$-spline basis of dimension $K=40$.

\begin{figure}
	\centering
	\includegraphics[width=\textwidth]{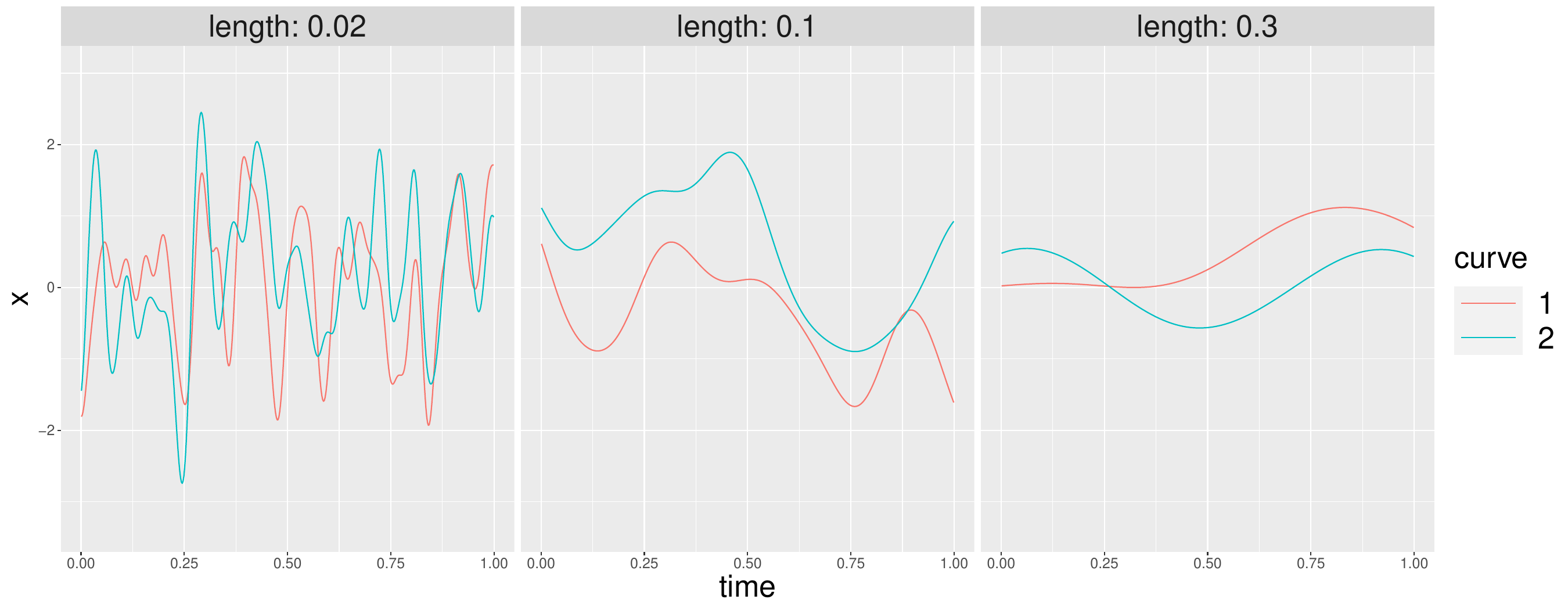}     
	\includegraphics[width=\textwidth]{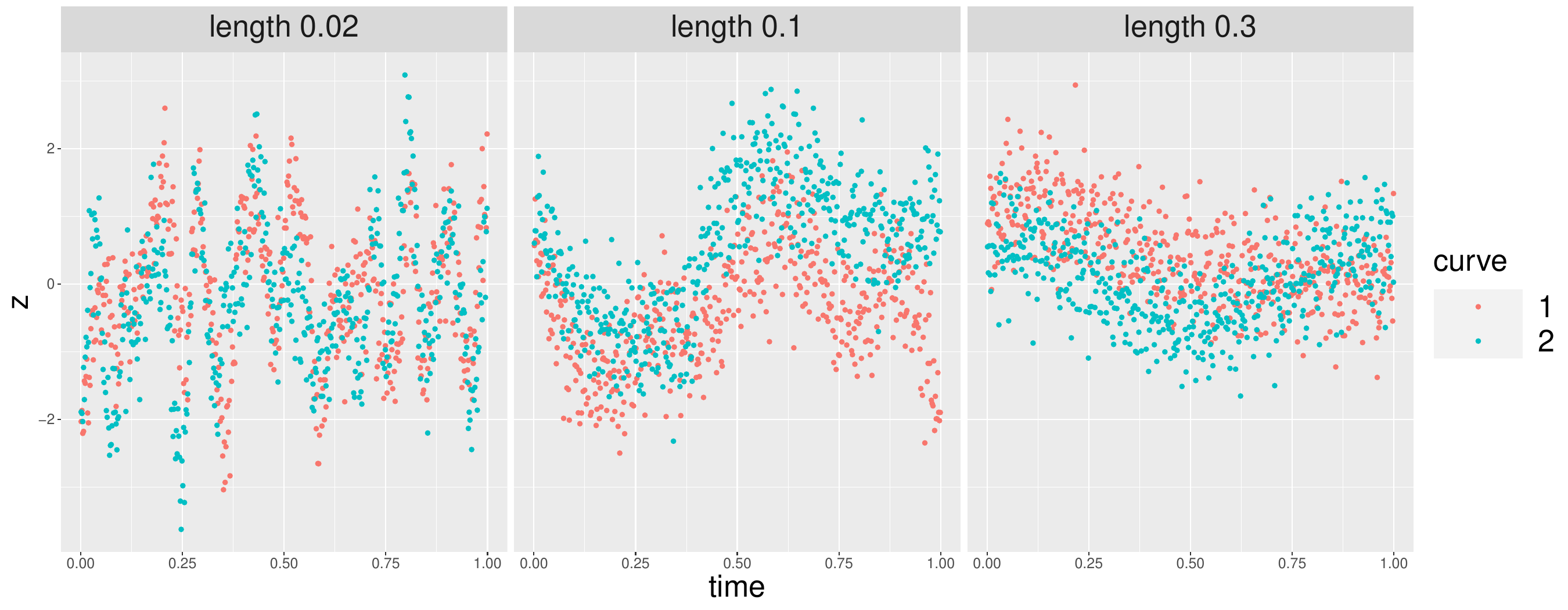}     
	\includegraphics[width=\textwidth]{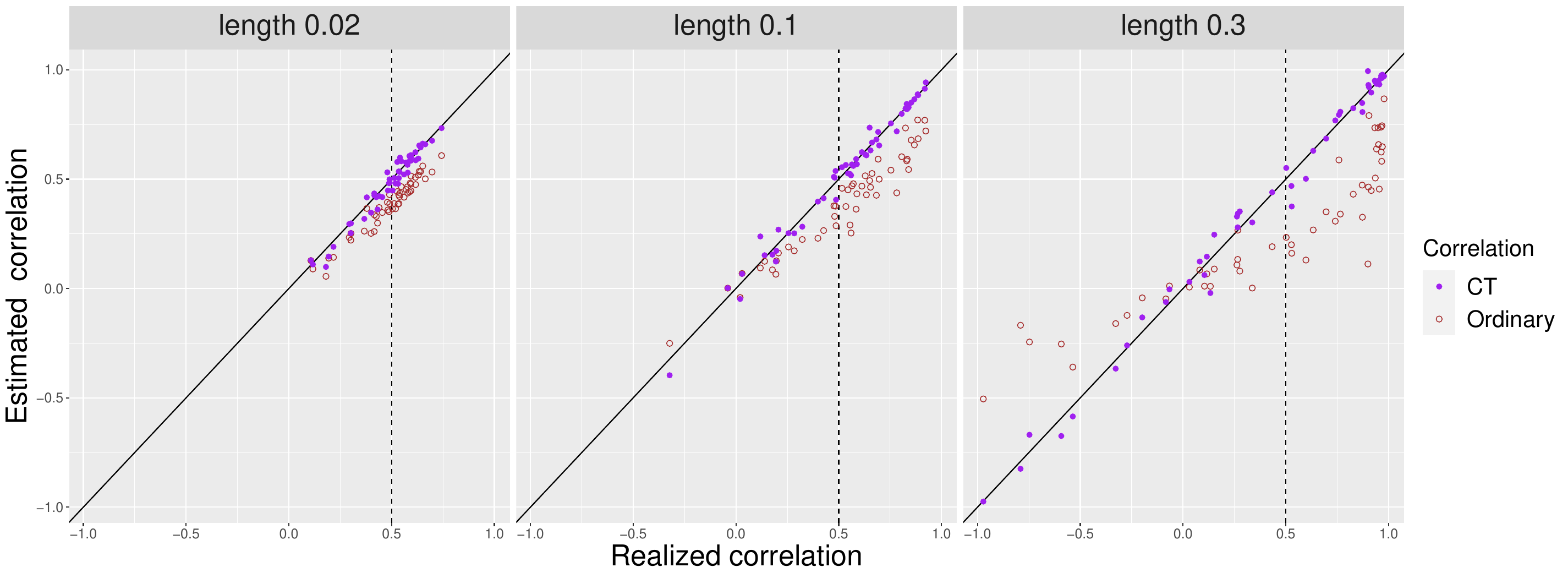}     
	\caption{Top row: Examples of bivariate Gaussian processes on $[0,1]$ with underlying between-curve correlation 0.5 and squared exponential within-curve correlation \eqref{squexp}, with length parameter $\ell=0.02, 0.1, 0.3$. Middle row: Noisily observed versions of the same example pairs of curves. Bottom row: Realized correlations $r^*$, and estimates $\hat{r},\hat{r}^*$ by ordinary and continuous-time correlation, for simulated bivariate processes, with each panel displaying results for 50 simulated noisy curve pairs akin to that shown in the panel directly above it. Solid line is the line of identity; dashed vertical line indicates the process correlation 0.5.}\label{biplabfig}
\end{figure}

\figref{biplabfig}  demonstrates how the CT correlation estimate $\hat{r}^*$ outperforms the ordinary sample correlation $\hat{r}$.  The top row displays example curve pairs $x_1(t),x_2(t)$ generated from a bivariate Gaussian process on $[0,1]$ with underlying between-curve   correlation 0.5 as above, and within-curve (auto)correlation \eqref{squexp} for length parameter $\ell=0.02, 0.1, 0.3$.   For each of the above three values of $\ell$, we generated 50 pairs of curves $x_1,x_2$, like those shown in the top row of \figref{biplabfig}, and for each pair we sampled noisy observations (as described in the previous paragraph), as displayed in the middle row. In  the bottom row we have plotted the realized correlations $r^*$ of the noiseless curves $x_1,x_2$, against the corresponding correlation estimates, $\hat{r}$ (ordinary) and $\hat{r}^*$ (continuous-time); each of the 50 points in each panel represents a pair of curves like those in the top row.

The horizontal spread of the points in the bottom row represents variation of the realized $r^*$ about the underlying process correlation 0.5 [step (i) of \eqref{3step}], which is seen to increase as the length parameter $\ell$ increases. Deviation of the points from the line of identity represents error in the estimates with respect to the realized $r^*$ [step (ii) of \eqref{3step}]. This error also increases with $\ell$, in particular for ordinary correlation, which is markedly biased toward zero. This bias is easy to explain: the noise term in \eqref{zij} inflates variance without affecting between-curve covariance, resulting in diminished correlation. Such attenuation of correlation due to noise is a long-established phenomenon \citep[e.g.,][]{spearman1910} and is sometimes corrected by multiplying the estimated correlation by the geometric mean of the two variables' reliabilities \citep{cohen2013}. 
For time series as opposed to independent observations, such a correction is not straightforward to implement, but the CT correlation approach proceeds by instead denoising the observed curves, resulting here in consistently accurate estimation of the between-curve correlation $r^*$. This simulation study, then, offers clear and concrete evidence for the benefits of CTMVA, even when ordinary MVA is feasible. Appendix~\ref{moresim} displays the results of (i) simulations as above but with underlying correlation 0.2 and 0.8, and (ii) simulations in which we vary the noise standard deviation $\sigma$ as well as $n$, the number of points at which the curves are observed. The overall conclusion is that unless $n$ is small and/or the curves are very bumpy, the CT correlation outperforms ordinary correlation, in some cases dramatically.

\section{Example: Temperature and air pollution in Chicago}\label{chisec}
\begin{figure}
	\begin{center}
		\includegraphics[width=.85\textwidth]{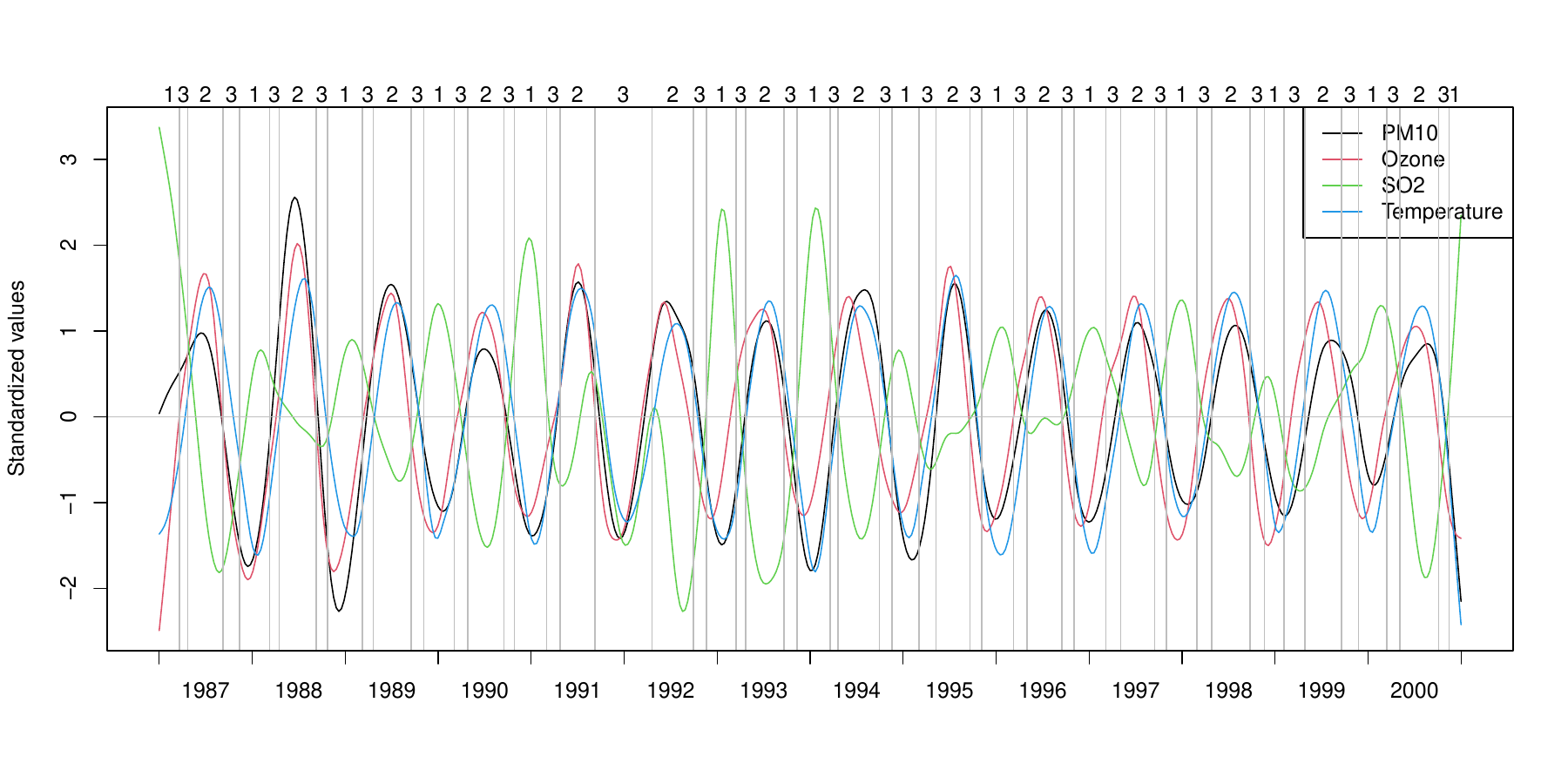}
		\includegraphics[width=.8\textwidth]{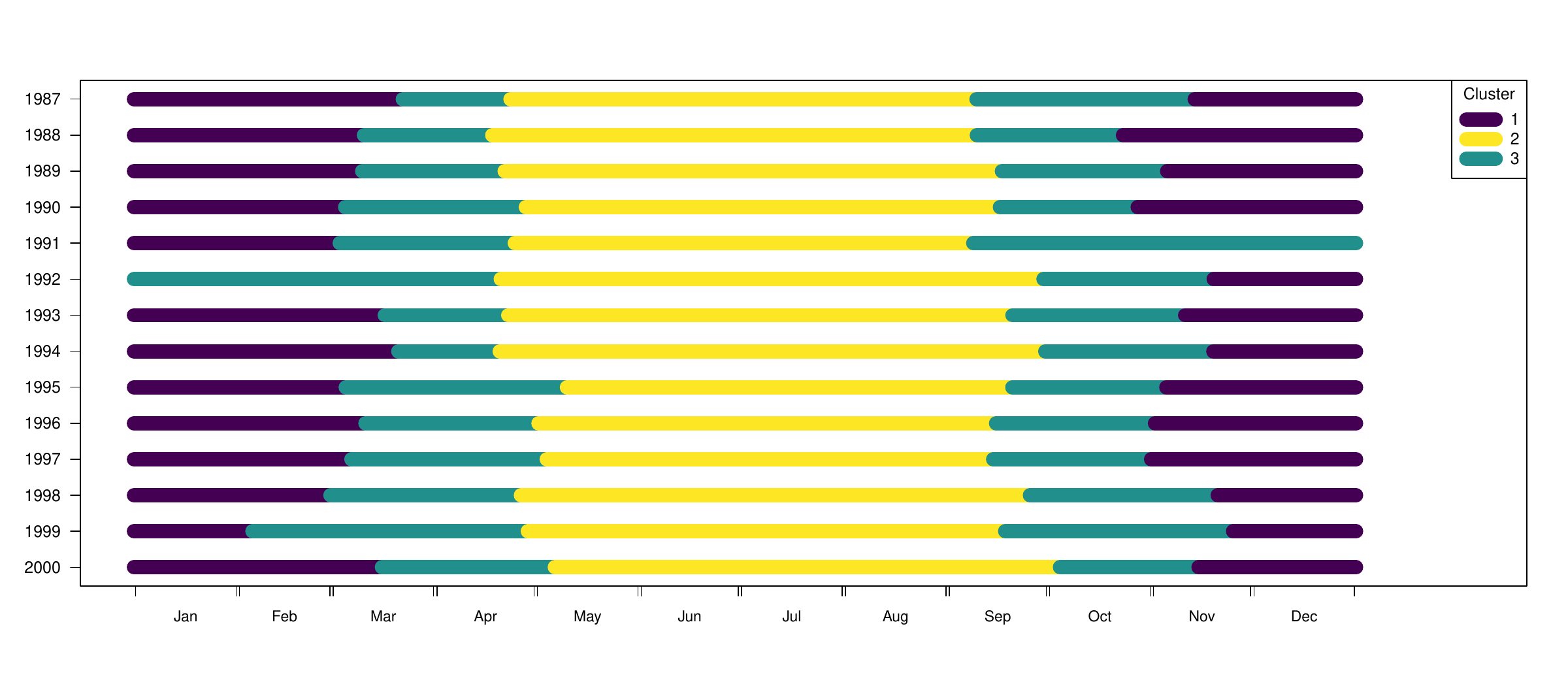}
		\includegraphics[width=.8\textwidth]{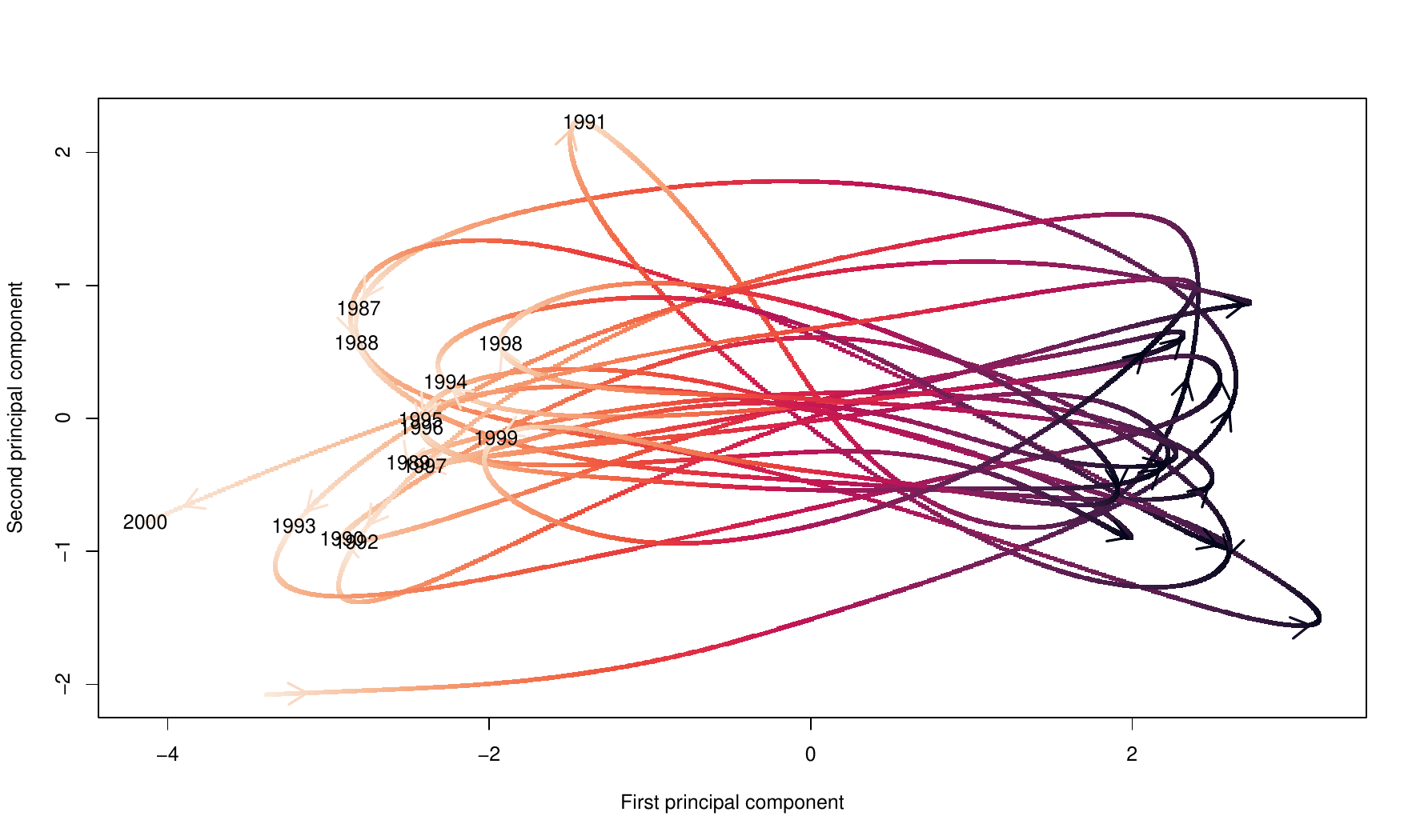}
	\end{center}
	\caption{Top: The four standardized and smoothed variables from the Chicago air pollution data. Vertical lines indicate transitions among the clusters found by CT 3-means clustering, with cluster numbers shown along the top of the plot. Middle: The same cluster transitions displayed separately by year. Bottom: Trajectory of the first two PCs of the Chicago air pollution data. Each label ($1987,\ldots,2000$) marks the end of the year indicated.}\label{chicagokmeans}
\end{figure}

\cite{wood2017} performed several analyses relating air pollution to death rates, using a data set of daily indicators for Chicago for the years 1987 to 2000 made available by \cite{peng2004}. Here we focus on the four predictors in these analyses, namely PM$_{10}$ (median 2.5- to 10-$\mu$m particles per m$^3$), ozone (in parts per billion), SO$_2$ (median sulfur dioxide measurement), and temperature. Our goal here is exploratory analyses of these four metrics using the CTMVA methods of \secref{pca}--\secref{kmeans}. All four variables were standardized and were smoothed using a 200-dimensional $B$-spline basis and residuals with an AR(1) residual structure \citep{wood2017}. 

\figref{chicagokmeans} displays results from continuous-time 3-means clustering. The top panel shows the four standardized variables' time series with vertical lines indicating transitions among the three clusters. As can be discerned more clearly in the middle panel, which shows the clustering separately by year, there is a fairly consistent seasonal pattern: cluster 1 corresponds roughly to the winter, cluster 2 to the summer, and cluster 3  to the spring and autumn. Thus each calendar year is divided into portions belonging to clusters 1, 3, 2, 3, 1, in that order, with one exception: the winter of 1991--92, in which cluster 1 does not appear. The clear seasonal pattern in the clusters seems to be driven largely by temperature variations, along with the high positive correlations of PM$_{10}$ and ozone with temperature.     SO$_2$, on the other hand, is negatively correlated with temperature. In the winter of 1991--92, however, SO$_2$ uncharacteristically dropped along with the other three variables, and this may help explain the non-occurrence of cluster~1. Appendix \ref{sil} provides further details of this analysis, in particular the silhouette width and the stability of the 3-means solution. We note that when ordinary 3-means clustering is applied to the raw time series, there are over 1100 transitions among the resulting three clusters, making it difficult at best to detect the cyclic seasonal pattern that emerges clearly from the continuous-time analysis.

The above results are complemented by CT principal component analysis, which found that most of the variation is explained by two PCs: 77.3\% by the first PC, with loadings 0.524, 0.505, -0.417, 0.544 for PM$_{10}$, ozone, SO$_2$ and temperature, respectively, and 15.0\% by the second PC, with loadings -0.351, -0.347, -0.870, -0.006. 
In view of the correlations noted above, the first PC is expected to be highest in the summer and lowest in the winter. This is confirmed by the bottom panel of \figref{chicagokmeans}, which displays
the trajectory of the first two PCs for the 14-year period, with labels ($1987,\ldots,2000$) marking the end of each year. The colors are lightest at the beginning and end of each year and darkest in the middle of the year, and this makes clear the annual oscillation of the first PC.  Both PCs' year-end scores are  highest for 1991, suggesting again that the winter of 1991--92 was atypical.

\begin{figure}
	\begin{center}
		\includegraphics[width=.39\textwidth]{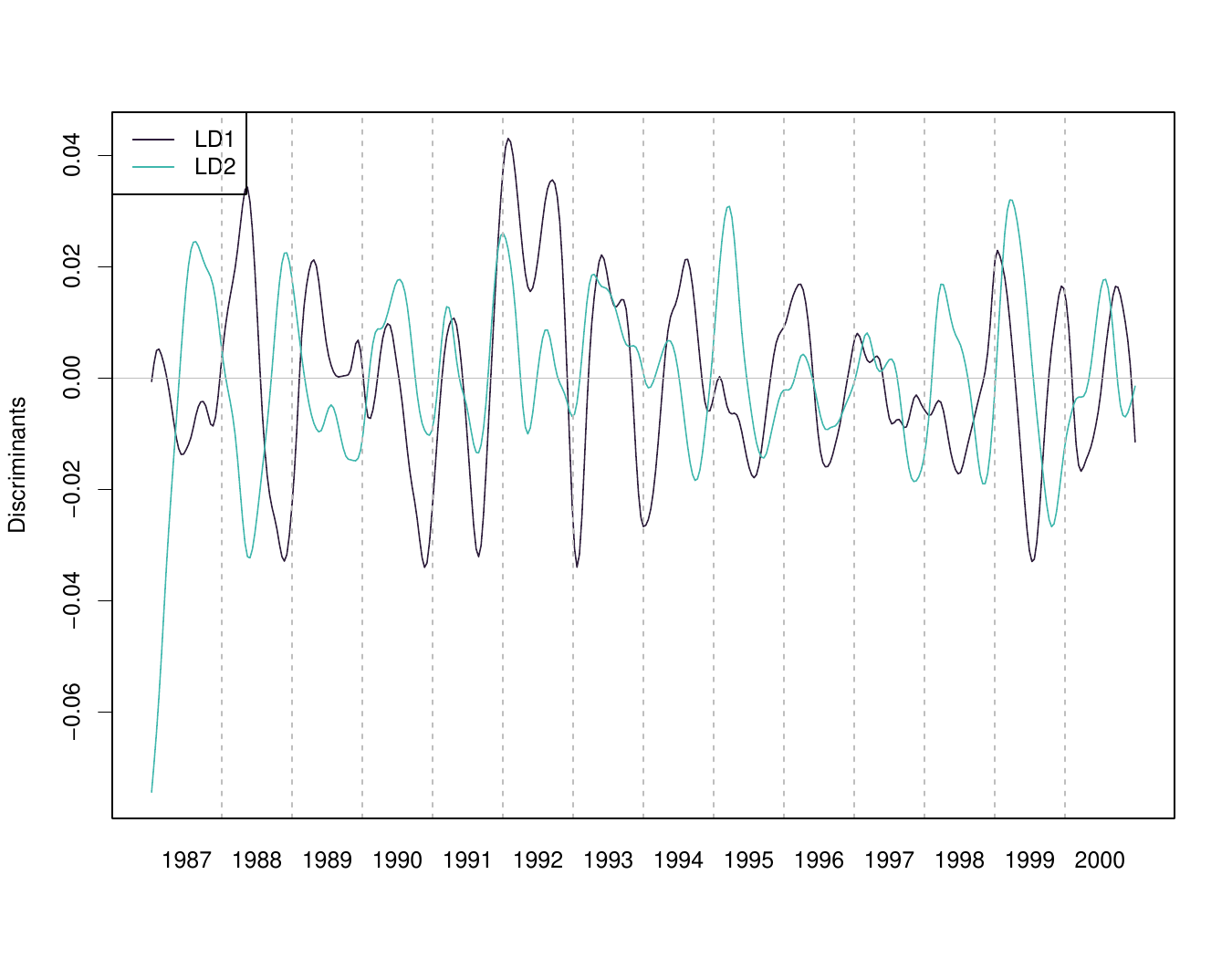}
		\includegraphics[width=.58\textwidth]{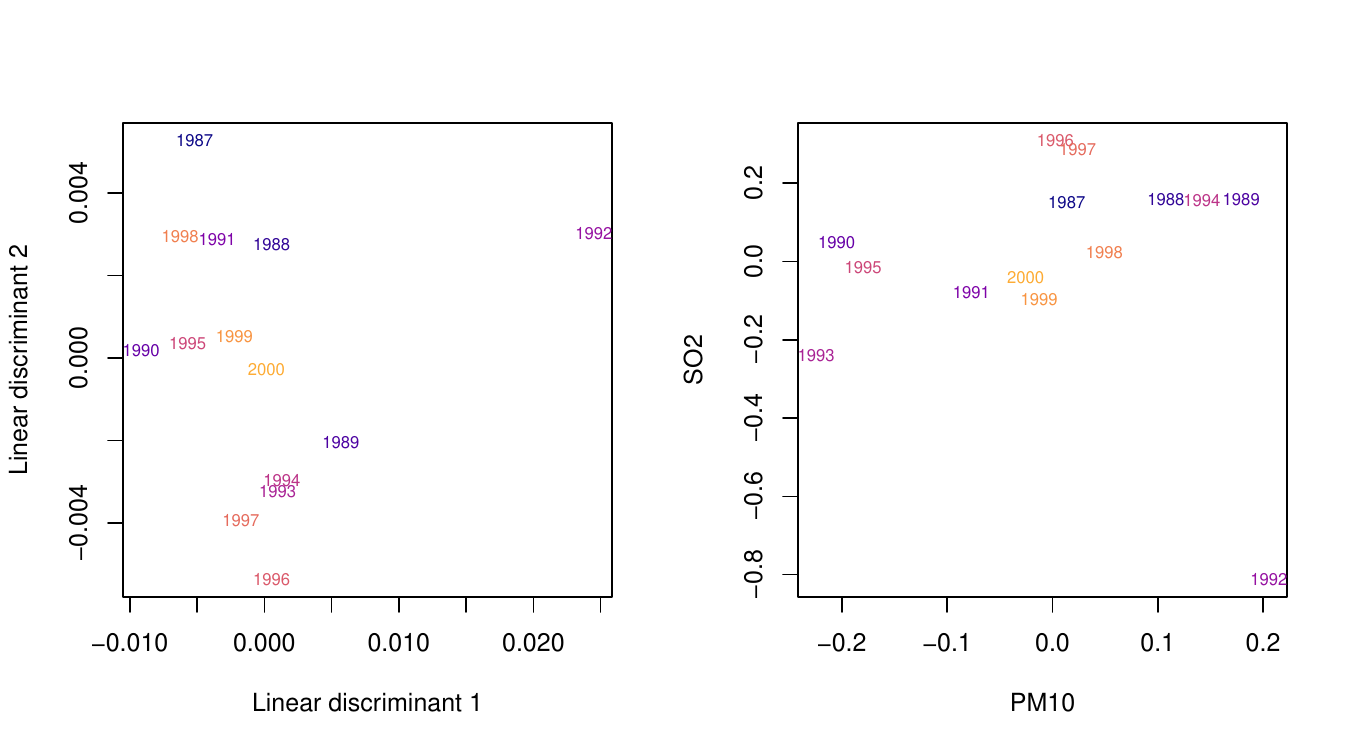}
	\end{center}
	\caption{Left: The first two linear discriminants for the Chicago air pollution data, displayed as functions of date. Center: Mean values of the first two discriminants for each of the 14 years. Right: Mean PM$_{10}$ and mean SO$_2$ for each year.}\label{ldplotz}
\end{figure}

Continuous-time Fisher's linear discriminant analysis, with the $G=14$ calendar years as subintervals among which we wish to discriminate, offers a way to compare and contrast the different years. 
The coefficients of PM$_{10}$, ozone, SO$_2$ and temperature, respectively, are given by $\bv_1=(0.0302, -0.0037, -0.0186,-0.0380)^T$ for the first discriminant and $\bv_2=(-0.0235,0.0134,  -0.0105 , 0.0034)^T$ for the second.   The plot of the time series of the first two discriminants in \figref{ldplotz} suggests that, as in the cluster analysis, the year 1992 stands out: the value of LD$_1$ tends to be very high during that year. This impression is confirmed by the middle panel, which shows the average values of the first two discriminants in each year and reveals that 1992 is a clear outlier for LD$_1$. This finding is explained by the very high mean PM$_{10}$ and extremely low mean SO$_2$  for 1992, as seen in the right panel.

\section{Discussion}\label{discuss}
Our simulation study in \secref{simsec} and 3-means cluster analysis in \secref{chisec} suggest that the denoised, smooth representation in CTMVA can, at least in some cases,  lead to improved performance over standard MVA with the raw data. In addition, as illustrated by the WDI example of \secref{wdisec},
CTMVA can be applied in certain settings for which classical MVA is unavailable, such as variables observed at  disparate times.
Another such setting is correlated point  processes, such as inhomogeneous Poisson processes.  \cite{ramsay2005} present a smooth basis function approach to estimating a time-varying intensity function $\iota(t)$. Given $p$ inhomogeneous Poisson processes with intensities $\iota_1(t),\ldots,\iota_p(t)$ on a common time interval, one can compute CT correlations among estimates of these intensities. However, more work is required on reliable smoothing parameter estimation for intensity functions.

While the applicability of CT covariance and correlation even for pairs of curves whose observation times have little or no overlap (as in the WDI example of \secref{wdisec}) is a clear advantage of the CT approach, there is an accompanying pitfall: as seen in the left panel of \figref{agofig}, curves are estimated less precisely within time ranges with fewer observations, and intuitively these times should be weighted less when estimating covariance. Formally, the covariance between the $u$th and $v$th curves would then be defined as 
$\int_{\cI}w(t)[x_u(t)-\bar{x}^*_u]][x_v(t)-\bar{x}^*_v]dt\mbox{ } / \int_{\cI}w(t)dt$ for some function $w(t)$  [generalizing \eqref{cts}]. Inverse variance is a standard choice for such a weight function, but it is not obvious how to incorporate the inverse variances of two function estimates into a single function $w(t)$. Minimizing the coefficient of variation, as in \cite{chen2014}, is a possible way forward.

A key objective of ongoing work is interval estimation for CT correlation, via analytic and resampling-based approaches. Optimal weighting, as described in the previous paragraph, may sometimes yield narrower confidence intervals.

Another aim of future work is to develop \emph{multilevel CTMVA} for multiple instances of continuous-time multivariate data, such as the full WDI and EEG data sets of \secref{wdisec} and  Appendix~\ref{eegsec}, respectively. This may be contrasted with two complex data structures extending the $n\times\infty$ setup of traditional FDA: \emph{multilevel} and \emph{multivariate} FDA. As a crude oversimplification, we may say that multilevel FDA \citep[e.g.][]{crainiceanu2009} considers $(n\times m)\times\infty$ data, in which a given function is observed $m$ times in each of $n$ individuals, whereas multivariate FDA \citep[e.g.][]{happ2018} concerns $n\times(v\times\infty)$ data, with $n$ instances of a collection of $v$ functions.   The multilevel CTMVA setup, of multiple instances of $\infty\times p$ data, differs from the either multilevel or multivariate FDA and may be more similar to the setup of \cite{dubin2005}. But whereas these authors estimate population dynamical correlation from a sample of $\infty\times p$-type observations, the aim of multilevel CT correlation would be to decompose the correlations among $p$ variables into ``within'' and ``between'' components.  Multilevel data will also allow for assessing the reliability of CTMVA results, via extensions of intraclass correlation \citep{xu2021}. 

Most of the methods of this paper are CT extensions of MVA methods involving $p\times p$ matrices. Certain other MVA methods, such as multidimensional scaling, begin with an $n\times n$ matrix, often representing dissimilarities or distances among $n$ observations, or alternatively similarities among them. Continuous-time ($n=\infty$) extensions are then more challenging than for methods based on  $p\times p$ matrices, since instead of $n\times n$ matrices we must work with their infinite-dimensional counterparts, namely bounded linear operators on $\cI$. We are pursuing CT methods of this type in ongoing work.

Our methods are implemented in the R package \texttt{ctmva} \citep{ctmva}, and code for our real-data analyses  can be found at \url{https://github.com/reissphil/ctmva}. 

%
%
%
%
%
%
%

\newpage
\appendix
\section*{APPENDICES}

\renewcommand{\thesection}{\Alph{section}}
\renewcommand{\thefigure}{A\arabic{figure}}
\renewcommand{\theequation}{A\arabic{equation}}

 \section{Daily average precipitation in Canada}\label{dapic}

\begin{figure}
	\centering
	\includegraphics[width=.89\textwidth]{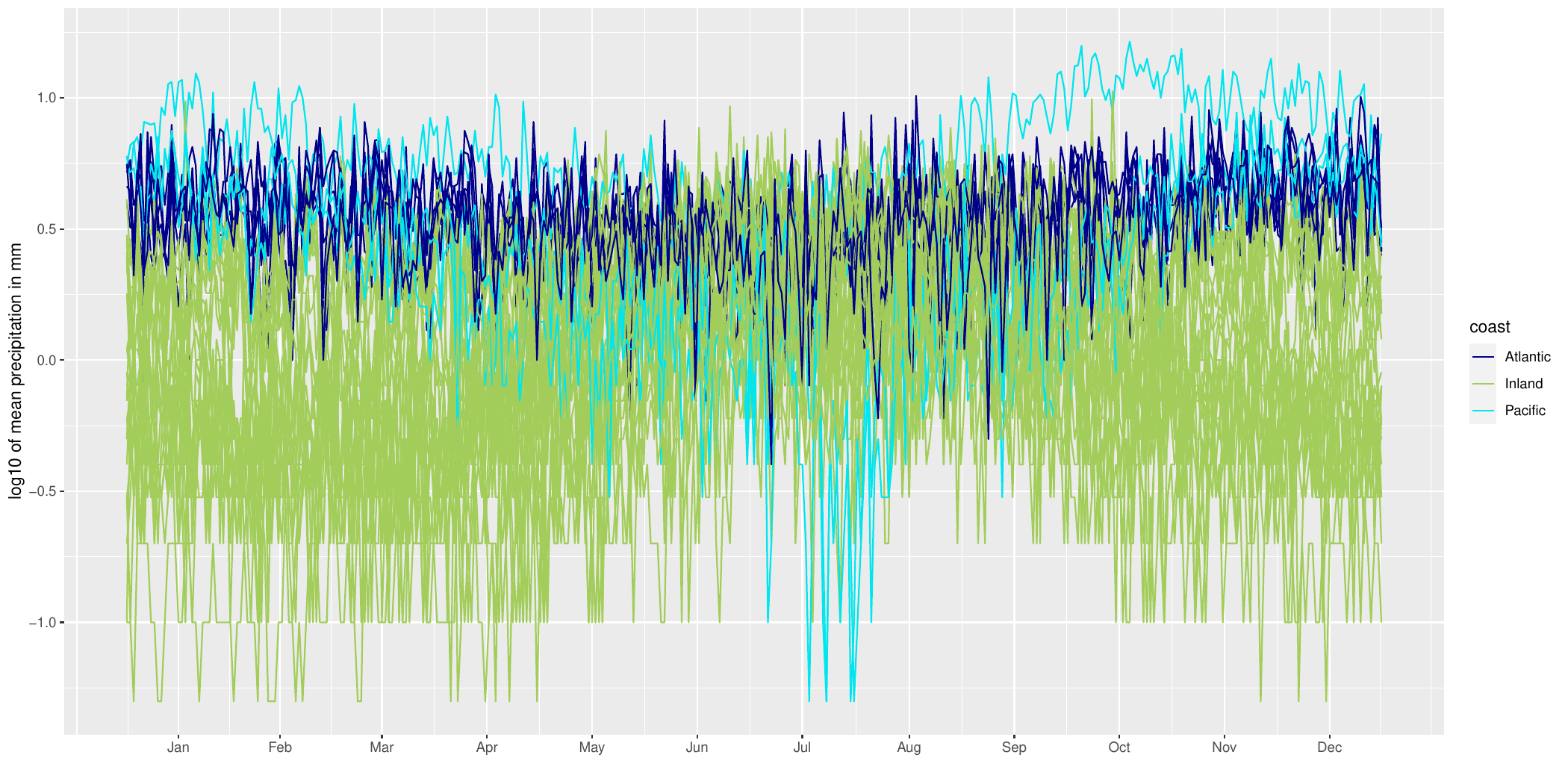}
	\includegraphics[width=.89\textwidth]{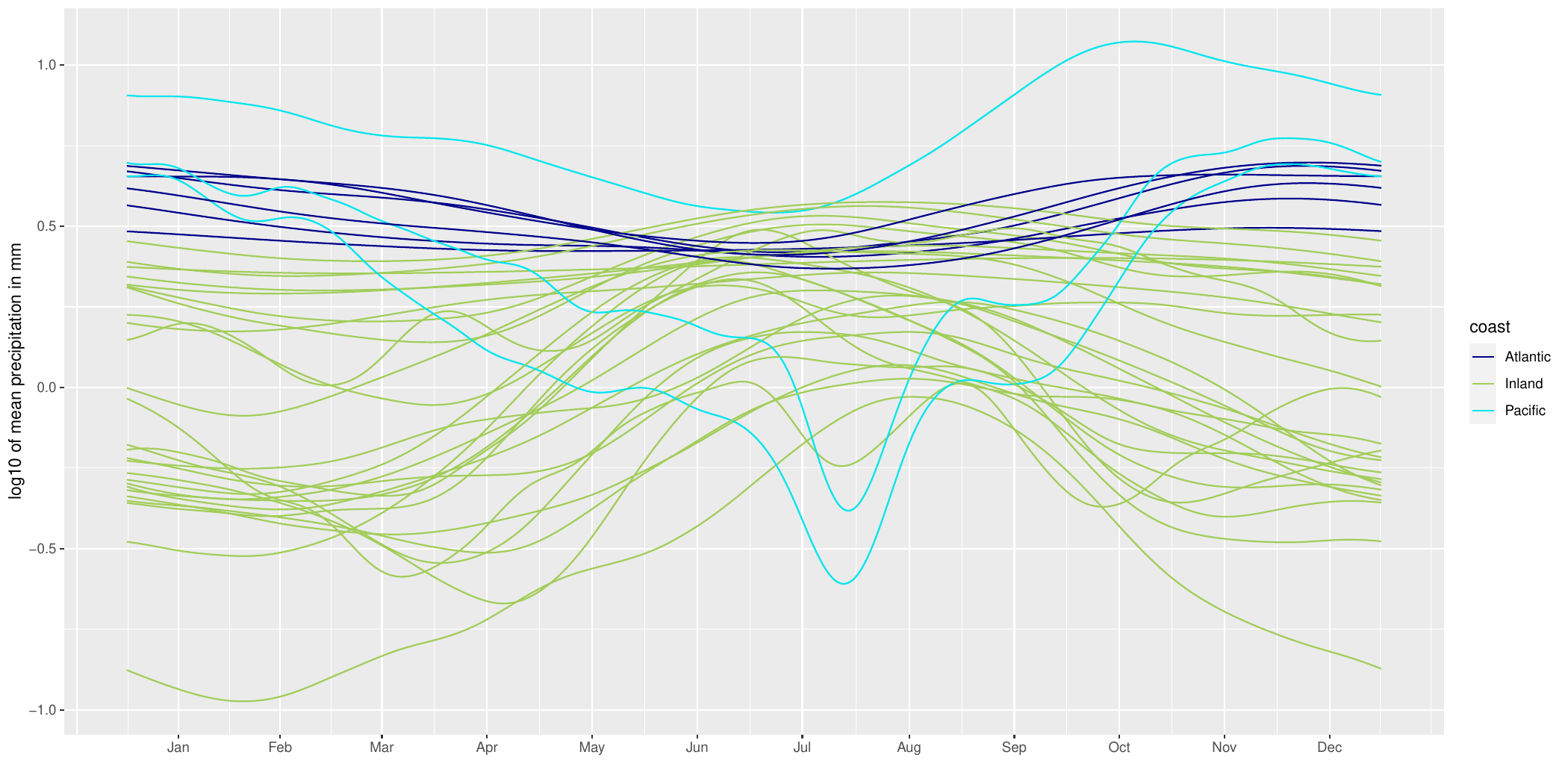}
	\includegraphics[width=.8\textwidth]{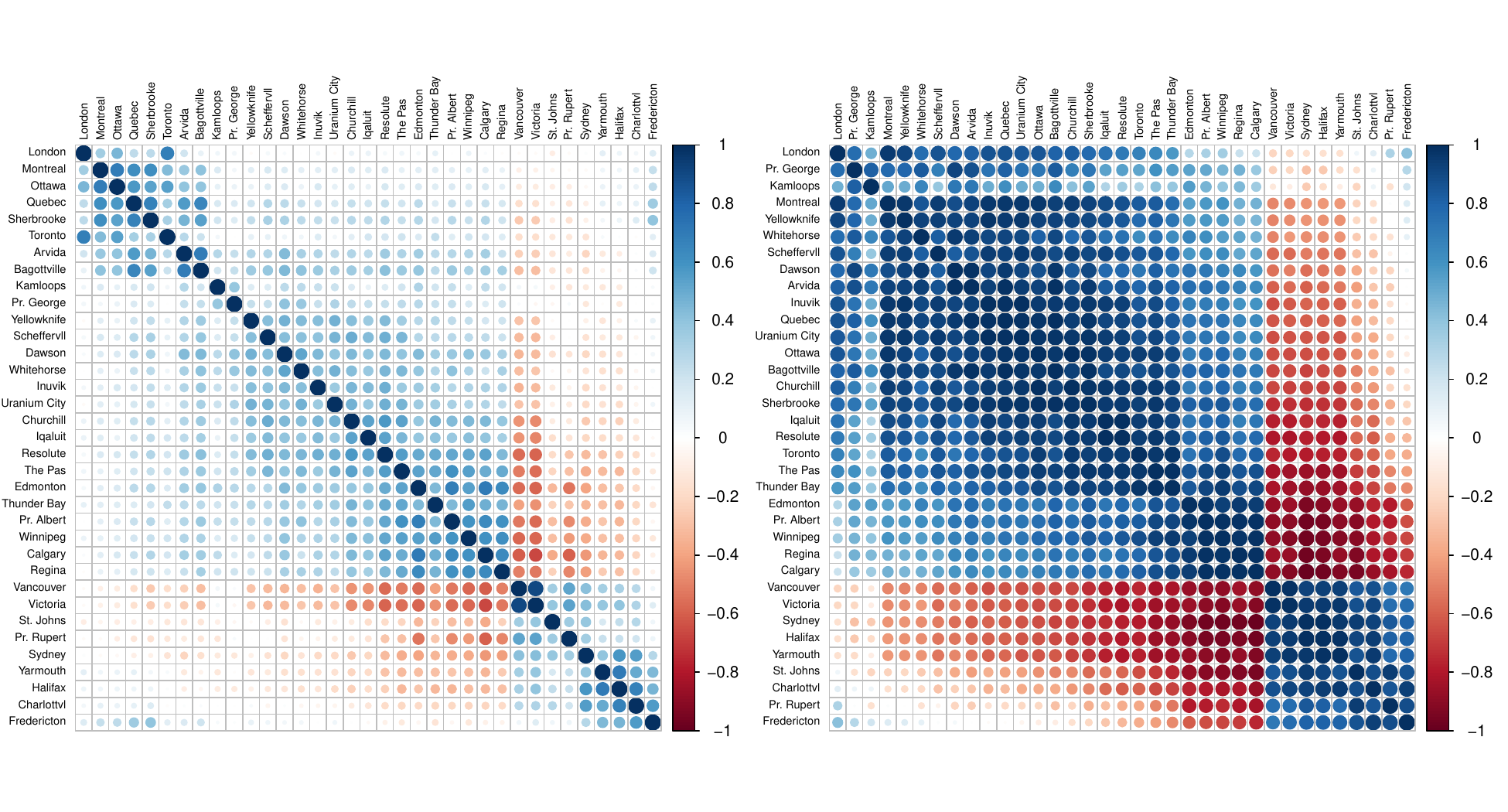}  \caption{Base-10 log mean daily precipitation in mm for the 35 Canadian weather stations of \cite{ramsay2005}, in raw form (above) and after smoothing (middle). Below: Correlations among  the 35 weather stations, obtained by treating the data as $n=365$ discrete observations (left) and by the continuous-time approach (right).}\label{preccor}
\end{figure}

The simulation study of \secref{simsec} demonstrated the contrast between ordinary versus CT correlations. As a real-data illustration of this phenomenon, we consider the Canadian precipitation data of \cite{ramsay2005}. As in \figref{tempcor}, the data consist of daily values for the 35 Canadian weather stations, but here the variable of interest is base-10 log of the average precipitation in millimeters \citep[see][]{fda}. \figref{preccor} displays the data in raw form (upper panel) and after smoothing (middle panel), using the same strategy as for the temperature data in \figref{tempcor}. The lower left panel of \figref{preccor} presents the ordinary Pearson correlation among the 35 stations' precipitation levels, when the data are treated as $n=365$ discrete observations of $p=35$ variables. 

The lower right panel shows the CT correlations based on the smooth curves. The CT correlations are much stronger than raw correlations, and the simulation results of \secref{simsec} support viewing the former as more accurate than the latter. 
For each correlation matrix, the stations are ordered by the angles formed by the leading two eigenvectors; this ordering helps to reveal patterns and groupings among the correlations \citep{friendly2003}.
In both correlation matrices the reordering puts the same nine stations at the end: the first six of the original ordering, which are located on the Atlantic coast, along  with Vancouver, Victoria and Prince Rupert, on the Pacific coast. The CT correlation matrix suggests that these nine stations form a tight cluster, with positive correlations among them and negative correlations between them and the inland stations. This pattern is much less clear in the ordinary correlation matrix, due perhaps to the attenuation phenomenon discussed in \secref{simsec}. As can be seen in the middle panel of \figref{preccor}, the Atlantic and Pacific stations tend to have the least precipitation in the summer, while the inland stations have an opposite pattern.

\begin{figure}
	\centering 
	\includegraphics[width=.9\textwidth]{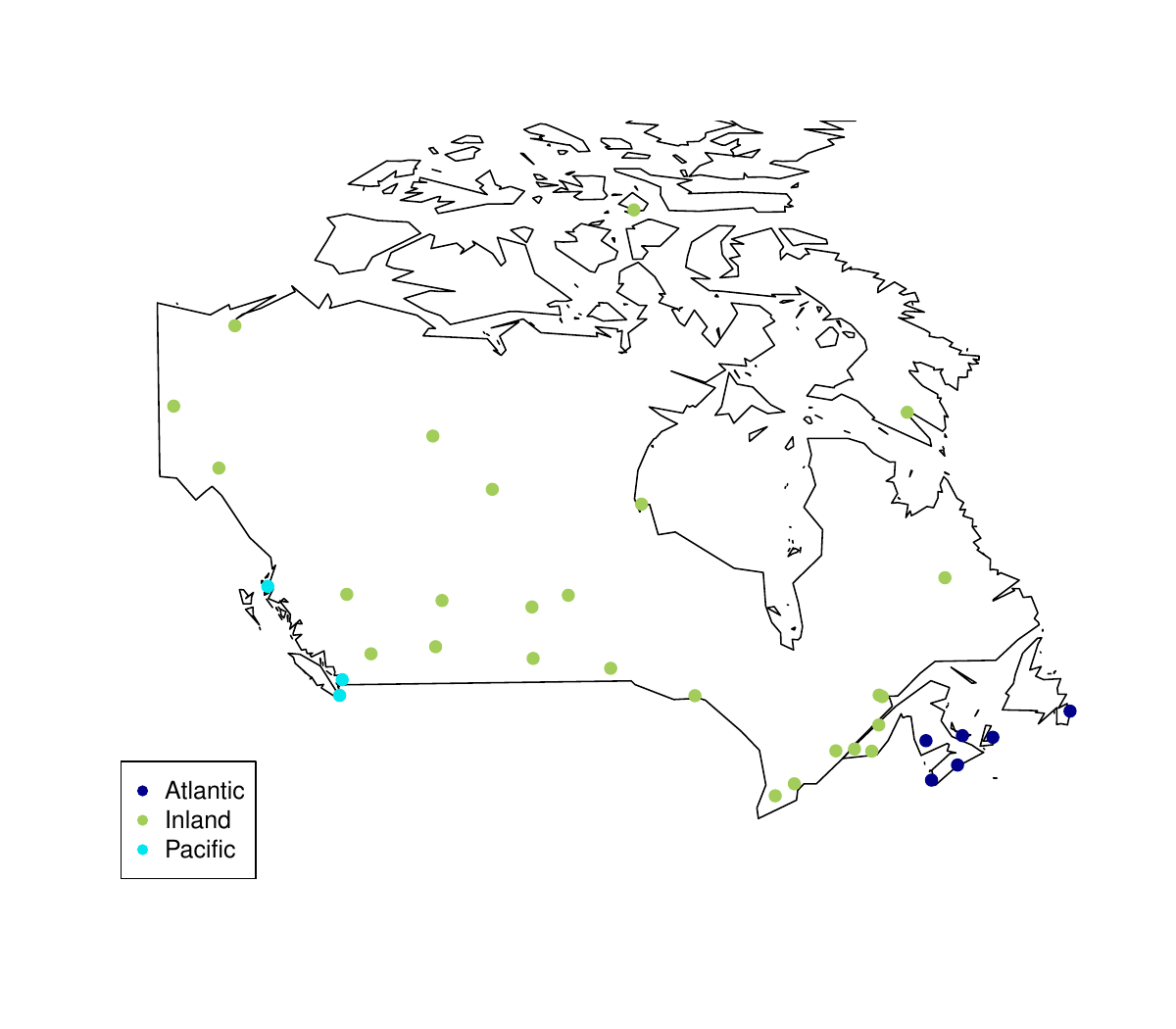}
	\caption{The 35 Canadian weather stations, color-coded by location on the Atlantic or Pacific coast  or inland.}\label{rrr}
\end{figure}

The CT correlation matrix of the log mean precipitation values for the 35 Canadian weather stations, displayed in \figref{preccor},    reveals that the nine coastal stations (six on the Atlantic  and three on the Pacific coast) tend to be highly correlated with each other and anticorrelated with the 26 inland stations.  \figref{rrr} provides a map of this division of the 35 weather stations into three subsets, using the same color-coding as in \figref{preccor}. We remark that \cite{ramsay2005}, as well as R package \texttt{fda} \citep{fda} and a number of subsequent analyses of the Canadian weather data, divide the stations into four regions, including 15 ``Atlantic'' and 5 ``Pacific'' stations. The nine stations referred to in the previous paragraph, and indicated in \figref{rrr}, meet a stricter criterion of proximity to the Atlantic or Pacific. 

\section{Computing \eqref{bmv}}\label{compdet}

For simplicity we consider only $\cS=\cI$. It is readily shown that the matrix $\bQ=|\cI|^{-1}\int_{\cI}[\bphi(t)-\bar{\bphi}][\bphi(t)-\bar{\bphi}]^Tdt$ of \eqref{bmv}
can be written as  \[\bQ=|\cI|^{-1}\bG-\bar{\bphi}\bar{\bphi}^T,\]
where $\bG=\int_{\cI}\bphi(t)\bphi(t)^Tdt$, i.e., the $K\times K$ matrix with $(k,\ell)$ element 
$g_{k\ell}=\int_{\cI}\phi_k(t)\phi_{\ell}(t)dt$. 

If $\bphi$ is an orthonormal Fourier basis on $\cI$ then $\bG=\bI_K$ and $\bar{\bphi}=(|\cI|^{-1/2},0,\ldots,0)^T$, so that $\bQ$ is simply the diagonal matrix with main diagonal $(0,|\cI|^{-1},\ldots,|\cI|^{-1})^T$.

For a $B$-spline basis,  $\bar{\bphi}$ and $\bG$ are obtained by numerical integration, with quadrature points $t_1,\ldots,t_Q$ and corresponding weights $w_1,\ldots,w_Q$. Let $\bPhi$ be the $Q\times K$ matrix of evaluation of the basis functions at the quadrature points, i.e., the $(q,k)$ entry of $\bPhi$ is $\phi_k(t_q)$. For basis objects in the R package \texttt{fda} \citep{fda}, this matrix can be generated with a call to function \texttt{eval.basis} from that package. Let $\bW$ be the $Q\times Q$ diagonal matrix whose diagonal elements are the quadrature weights. Then for $k=1,\ldots,K$,
$\bPhi^T\bW\bone_Q$ has $k$th element $\sum_{q=1}^Qw_q\phi_k(t_q)\approx\int_\cI\phi_k(t)dt$ and thus 
\[\bar{\bphi}\approx|\cI|^{-1}\bPhi^T\bW\bone_Q.\]
Similarly, $\bPhi^T\bW\bPhi$ has $(k,\ell)$ element $\sum_{q=1}^Qw_q\phi_k(t_q)\phi_{\ell}(t_q)\approx\int_{\cI}\phi_k(t)\phi_{\ell}(t)dt$,
and thus 
\[\bG\approx\bPhi^T\bW\bPhi.\]
Our implementation uses the the Newton-Cotes 7-point rule for numerical quadrature, as suggested by \cite{wand2008}. For cubic or lower-order $B$-splines, the Newton-Cotes 7-point rule makes the above integrals exact rather than approximate, so that $\bQ$ is obtained exactly.

\section{Diagnosing the need for detrending}\label{diagnose}
We present here a simple diagnostic to indicate, for a given data set, whether removing a common trend before computing the CT covariance or correlation (see \secref{2forms}) would be beneficial. If we adopt an FDA viewpoint toward the $p$ functions $x_1(t),\ldots,x_p(t)$, a strong common trend implies that all $p$ functions are quite similar to an underlying mean function $\mu_x(t)$. This similarity can be quantified by using the function \texttt{pffr} \citep{ivanescu2015,scheipl2015} from the R package  \texttt{refund} \citep{refund} to fit the simplest possible functional-response model, namely $x_i(t)=\mu_x(t)+\varepsilon_i(t)$ for $i=1,\ldots,p$. Here $\mu_x(t)$ is a ``functional intercept,'' in line with the syntax for this model, \texttt{pffr(X $\sim$ 1)} where \texttt{X} is a $p$-row matrix containing the functions. The (adjusted) $R^2$ outputted by \texttt{pffr} for this model \citep[based on the underlying model fitted by the R package \texttt{mgcv} of][]{wood2017} gives the variance explained, for the collection of all points $x_i(t)$ in all $p$ functions, by the fitted values $\hat{x}_i(t)=\hat{\mu}_x(t)$. This ``pointwise'' $R^2$ \citep[as opposed to the \emph{functional} $R^2$ of][which equals zero for this model]{muller2008} provides a measure of the strength of the common trend. For the Canadian weather data set, $R^2$ is found to be 71\% for the temperature curves, but only 4.9\% for the log precipitation curves; hence our decision to detrend the former (see \secref{2forms}) but not the latter (see Appendix \ref{dapic}).

\section{Electroencephalography data}\label{eegsec}
\begin{figure} \centering\includegraphics[width=.9\textwidth]{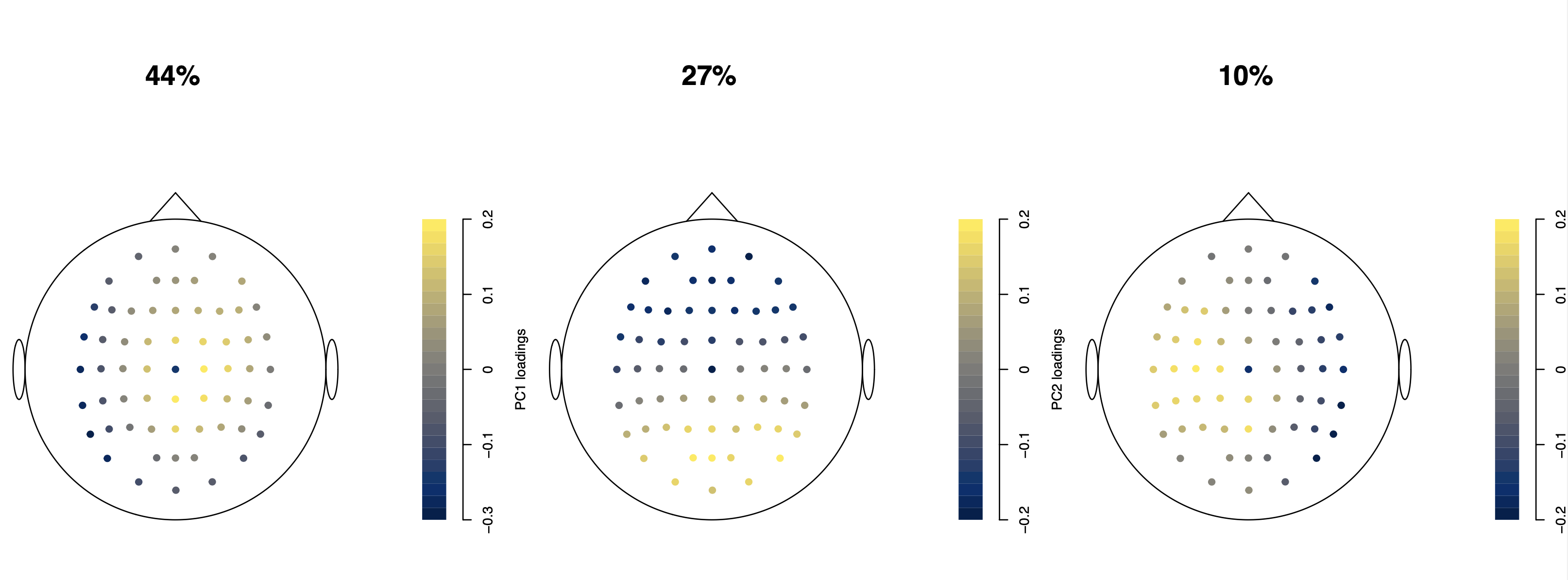}  
	\caption{Loadings of the first three continuous-time principal components for an example EEG trial. panel titles give the percent variance explained by each PC.}\label{eegfig}
\end{figure}

We applied continuous-time PCA  to a portion of the Begleiter electroen\-ceph\-a\-log\-ra\-phy (EEG) data set from the UCI Machine Learning Repository, made available in R by \cite{eegkit}. The data consist of multiple EEG trials for each of a set of individuals, with each trial consisting of electrical signals at $p=64$ channels with 256 time points within 1 second.
Loadings for the first 3 PCs for a randomly selected trial are shown in \figref{eegfig}, and these have reasonably clear interpretations: PC1 roughly captures the balance between signal in brain regions that are more central versus more peripheral (as viewed from above), PC2 represents anterior versus posterior regions, and PC3 represents left versus right hemisphere.

As suggested in \secref{discuss}, a natural next step would be multilevel CT-PCA to analyze multiple trials nested within individuals, as are available with this data set. The aim would be similar to the multilevel, longitudinal and structured variants of functional PCA \citep{di2009,greven2010,shou2015,cui2023}, but rather than decom\-pos\-ing a covariance operator, the structure being decomposed is a $p\times p$ covariance or correlation matrix as in \cite{timmerman2002}.   We leave this development to future work. 

\section{Continuous-time canonical correlation analysis}\label{cca}

In classical canonical correlation analysis (CCA)\ we are given an $n\times p$ matrix $\bX$ and an $n\times q$ matrix $\bY$, and we seek \emph{pairs} of unit vectors $\ba\in\mathbb{R}^p,\bb\in\mathbb{R}^q$ such that the sample correlation of $\ba^T\bx,\bb^T\by$ is maximized in an iterative sense:
\begin{enumerate}
	\item $\ba_1^T\bx,\bb_1^T\by$ attain the maximal sample correlation among all pairs $(\ba,\bb)$ of unit vectors $\ba\in\mathbb{R}^p,\bb\in\mathbb{R}^q$.
	\item For $m=2,\ldots,\min\{p,q\}$, $\ba_m^T\bx,\bb_m^T\by$ attain the maximal sample correlation among all pairs $(\ba,\bb)$ of unit vectors such that $\ba^T\bx$ is uncorrelated with $\ba_1^T\bx,\ldots,\ba_{m-1}^T\bx$ and $\bb^T\by$ is uncorrelated with $\bb_1^T\by,\ldots,\bb_{m-1}^T\by$.
\end{enumerate}
Then $(\ba_m^T\bx,\bb_m^T\by)$ are the $m$th pair of canonical variables, and their correlation is the $m$th canonical correlation.

The solution to the iterative maximization problem is given in terms of the covariance matrices of the two sets of variables, as well as their cross-covariance matrix: 
\[\bS_{xx}=n^{-1}\bX^{cT}\bX^c,\quad\bS_{yy}=n^{-1}\bY^{cT}\bY^c,\quad\bS_{xy}=n^{-1}\bX^{cT}\bY^c,\quad\bS_{yx}=\bS_{xy}^T,\]
where $\bX^c,\bY^c$ are column-centered versions of the two matrices and $\bS_{xx},\bS_{xy}$ are assumed to be nonsingular.
It can be shown \citep[e.g.,][]{rencher2012} that 
\[\bM_1=\bS^{-1}_{xx}\bS_{xy}\bS^{-1}_{yy}\bS_{yx}\quad\mbox{and}\quad\bM_2=\bS^{-1}_{yy}\bS_{yx}\bS^{-1}_{xx}\bS_{xy}\]
have the same positive eigenvalues $r_1^2\geq r_2^2\geq\ldots\geq r^2_{\min\{p,q\}}$, and that if $\ba_i,\bb_i$ are eigenvectors of $\bM_1,\bM_2$ respectively corresponding to the $i$th largest eigenvector, then $\ba_i^T\bx$, $\bb_i^T\by$ are the $i$th canonical variates, with canonical correlation $r_i$, which can be forced to be positive by multiplying either $\ba_i$ or $\bb_i$ by -1 if needed.

To develop a continuous-time version of CCA, suppose that  two sets of functions, $\bx(t)=[x_1(t),\ldots,x_p(t)]^T$ and $\by(t)=[y_1(t),\ldots,y_q(t)]^T$, are defined on $\cI$. 
We now seek pairs $(\ba,\bb)$ of unit vectors defining linear combinations $\ba^T\bx(t),\bb^T\by(t)$ of the two sets of functions that have maximal correlation in the above iterative sense, but here, generalizing \eqref{cts}, the covariance of $\ba^T\bx(t),\bb^T\by(t)$ is given by 
\[|\cI|^{-1}\int_\cI [\ba^T\{\bx(t)-\bar{\bx}\}] [\bb^T\{\by(t)-\bar{\by}\}]dt,\]
and their correlation is then given by
\[\frac{\int_\cI [\ba^T\{\bx(t)-\bar{\bx}\}] [\bb^T\{\by(t)-\bar{\by}\}]dt}{\sqrt{\int_\cI [\ba^T\{\bx(t)-\bar{\bx}\}]^2dt\int_\cI [\bb^T\{\by(t)-\bar{\by}\}]^2dt}}.\]
These pairs of linear combinations may be called the \emph{canonical functions}.

We assume for simplicity  that both sets of functions are represented with respect to a common $K$-dimensional basis $\bphi$, i.e.,
\[\bx(t)=\bC^T_x\bphi(t),\quad\by(t)=\bC^T_y\bphi(t)\]
for some matrices $\bC_x,\bC_y$ of dimension $K\times p$, $K\times q$ respectively.
As with PCA, the continuous-time solution to the iterative maximization problem ensues upon ``translating'' the relevant matrices to the continuous-time setting. We do so by defining
\[\bM^*_1=\bS^{*-1}_{xx}\bS^*_{xy}\bS^{*-1}_{yy}\bS^*_{yx}\quad\mbox{and}\quad\bM^*_2=\bS^{*-1}_{yy}\bS^*_{yx}\bS^{*-1}_{xx}\bS^*_{xy},\]
where 
\[\bS^*_{xx}=\bC^T_x\bQ\bC_x,\quad\bS^*_{yy}=\bC^T_y\bQ\bC_y,\quad\bS^*_{xy}=\bC^T_x\bQ\bC_y^T,\quad\bS^*_{yx}=\bS_{xy}^{*T},\]
with $\bQ$ defined as in \eqref{bmv}. Then, as in the classical case, $\bM^*_1,\bM^*_2$ have the same positive eigenvalues; and if $\ba_i,\bb_i$ are eigenvectors of $\bM^*_1,\bM^*_2$ respectively corresponding to the $i$th largest eigenvector, then $\ba_i^T\bx(t)$, $\bb_i^T\by(t)$ are the $i$th canonical functions.

\section{Classical Fisher's linear discriminant analysis, and the CT within-between decomposition}\label{flda}
As noted in \secref{lda}, classical Fisher's linear discriminant analysis (LDA) seeks linear combinations $\bv^T\bx$ that optimally separate $G\geq 2$ \emph{a priori} subsets $\bX_1,\ldots,\bX_G$  of the data set.
Assume that, for $g=1,\ldots,G$, $\bX_g$ is $n_g\times p$ with $i$th row $\bx_{gi}^T$; let 
\[\bX=\left[\begin{array}{c}\bX_1\\\vdots\\\bX_G\end{array}\right]\]
be the $n\times p$ combined data set, where $n=n_1+\ldots+n_G$; and let $\bar{\bx}=n^{-1}\bX^T\bone_n$ and $\bar{\bx}_g=n_g^{-1}\bX_g^T\bone_{n_g}$ denote the overall and group-$g$ mean vectors. We proceed by partitioning the total sums of squares and cross-products matrix $\bT=n\bS=\sum_{g=1}^G\sum_{i=1}^{n_g}(\bx_{gi}-\bar{\bx})(\bx_{gi}-\bar{\bx})^T$ into \emph{within-groups} and \emph{between-groups} components, $\bT=\bW+\bB$,
where 
\[\bW= \sum_{g=1}^G\sum_{i=1}^{n_g}(\bx_{gi}-\bar{\bx}_g)(\bx_{gi}-\bar{\bx}_g)^T,\quad
\bB = \sum_{g=1}^Gn_g(\bar{\bx}_g-\bar{\bx})(\bar{\bx}_g-\bar{\bx})^T.\]
With these definitions, $\bv^T\bW\bv$ is the sum of squared deviations of the linear combinations $\bv^T\bx_{gi}$ from their within-group means, while $\bv^T\bB\bv$ is a weighted sum of these within-group means from the overall mean, with weight $n_g$ for group $g$. (Some authors formulate Fisher's LDA differently, with these weights omitted. Moreover, some authors divide $\bW$ and $\bB$ by constants that allow them to be interpreted as within- and between-group covariance matrices, but this does not change the solution.) The linear combination $\bv_1^T\bx$ where $\bv_1=\argmax_v \bv^T\bB\bv/\bv^T\bW\bv$ can be said to maximally separate the groups, relative to the within-group variation. It can be shown that  $\bv_1$ is the eigenvector of $\bW^{-1}\bB$ corresponding to its largest eigenvalue. If $s$ $(\leq\min\{G-1,p\})$ is the number of positive eigenvalues of $\bW^{-1}\bB$, then for $m=2,\ldots,s$, $\bv_m$, the maximizer of $\bv^T\bB\bv/\bv^T\bW\bv$ subject to $\bv^T\bW\bv_1=\ldots=\bv^T\bW\bv_{m-1}=0$,  is  given by the eigenvector of $\bW^{-1}\bB$ corresponding to its $m$th-largest eigenvalue. The linear combinations $\bv_1^T\bx,\ldots,\bv_s^T\bx$, which optimally separate the groups in the above iterative sense, are known as Fisher's linear discriminants. 

Similarly to the derivations of $\bar{\bx}^*,\bS^*$ in \secref{defsec},  the matrices $\bT^*,\bW^*,\bB^*$ in \secref{lda} can be derived as limits of their classical counterparts $\bT,\bW,\bB$ times $|\cI|/n$  for observations on a uniform grid of $n$ time points, as $n\rightarrow\infty$. Thus the CT decomposition $\bT^*=\bW^*+\bB^*$ can be inferred from the classical decomposition $\bT=\bW+\bB$. A more direct proof of the CT decomposition is as follows:
\begin{eqnarray*}
	\bT^*&=&\sum_{g=1}^G \int_{\cI_g}[\bx(t)-\bar{\bx}^*] [\bx(t)-\bar{\bx}^*]^Tdt \\
	&=&\sum_{g=1}^G \int_{\cI_g}[\{\bx(t)-\bar{\bx}^*_g\}+\{\bar{\bx}^*_g-\bar{\bx}^*\}] [\{\bx(t)-\bar{\bx}^*_g\}+\{\bar{\bx}^*_g-\bar{\bx}^*]\}^Tdt \\
	&=&\bW^*+\bB^*+2\sum_{g=1}^G[\bar{\bx}^*_g-\bar{\bx}^*]\int_{\cI_g}[\bx(t)-\bar{\bx}^*_g]dt.
\end{eqnarray*}
Since the integral in the last line equals zero by the definition of $\bar{\bx}^*_g$, we conclude that $\bT^*=\bW^*+\bB^*$ as claimed.

 \section{Computational aspects of CT $k$-means clustering}\label{kfast}
If the basis $\bphi(t)$ consists of cubic splines then, within the $j$th inter-knot interval \citep[for $j=1,\ldots,K-3$; see][]{ramsay2005}, $\bphi(t)^T=(1\quad t\quad t^2\quad t^3)\bL_j$ for a $4\times K$ matrix $\bL_j$ of polynomial coefficients. 
Substituting this into \eqref{a1a2}, we conclude that $A_1(t)-A_2(t)=(1\quad t\quad t^2\quad t^3)\bv_j$ where \[\bv_j=2\bL_j\bC(\bm_2-\bm_1)+\left[\begin{array}{c}\bm_1^T\bm_1-\bm_2^T\bm_2 \\ 0 \\0 \\0\end{array}\right].\]
Consequently, the zeroes of the cubic polynomial \eqref{a1a2} in the $j$th inter-knot interval can be found by inputting this vector $\bv_j$ to R function \texttt{polyroot}, which implements the Jenkins-Traub zero-finding algorithm. 

Once the data have been reduced to a spline representation, the computing time for CT $k$-means is insensitive to the number $n$ of observations in the raw data. In experiments with 20 spline basis functions and $p=4$ (using the same laptop that was used for the WDI analysis in \secref{wdisec}), it took about 5 seconds to run CT $k$-means for $k=2,3,\ldots,15$. By contrast, for ordinary $k$-means with the same range of values of $k$, as implemented by the standard R function \texttt{kmeans}, we found the computing time for the default algorithm of \cite{hartigan1979} to increase slightly faster than linearly with $n$, reaching about 80 seconds when $n\approx 3\times 10^6$. Moreover, in many instances, convergence was not attained at the ``quick-transfer'' stage \citep[step 6 in][]{hartigan1979}. Computing times were slightly lower with two less refined algorithms implemented in \texttt{kmeans}, one due to \cite{lloyd1982} and \cite{forgy1965} and the other due to \cite{macqueen1967}, but these often struggled to converge. It must be acknowledged that for large $n$ the time required to smooth the data with respect to the basis may at least partly offset the computation-time advantage of the CT method. This depends, however, on how one chooses the smoothing parameter in penalized basis smoothing. The R package \texttt{fda} \citep{fda} offers the function \texttt{Data2fd} with a simple default choice of the smoothing parameter. \cite{wood2017} summarizes some techniques for optimal smoothness selection with large $n$, while \cite{li2024} propose a faster technique suitable for even larger $n$.

\section{Silhouette width and stability of the 3-means analysis}\label{sil}
\begin{figure}
	\centering
	\includegraphics[width=\textwidth]{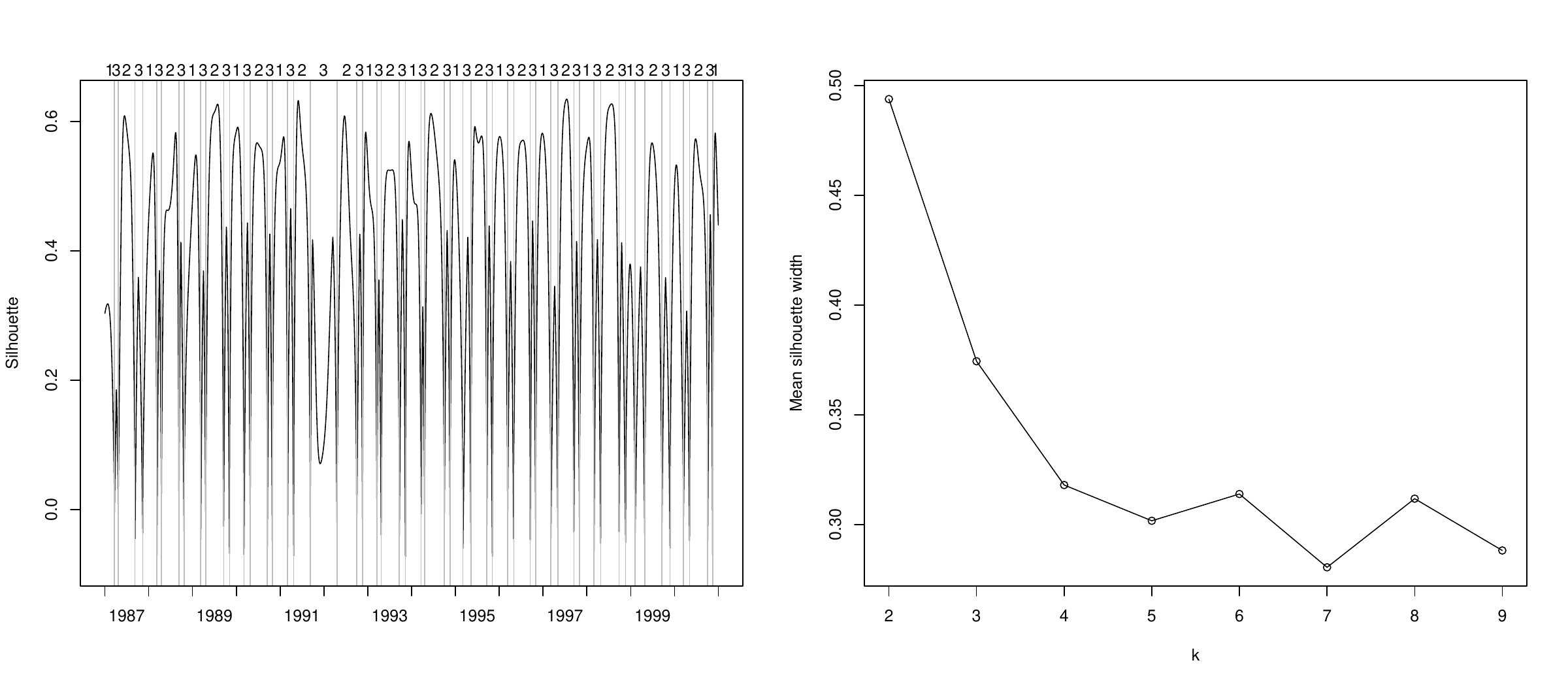}     
	\caption{At left, a plot of the silhouette $s(t)$ for the CT 3-means clustering of the Chicago air pollution data presented in  \secref{chisec}. At right, average silhouette width of $k$-means clusterings of the data, for $k=2,3,\ldots,9$.}\label{silfig}
\end{figure}

Given a clustering of $n$ discrete observations into $k>1$ clusters, \cite{rousseeuw1987} defines the silhouette, or silhouette width, of the $i$th observation as $s(i)=\frac{b(i)-a(i)}{\max\{a(i),b(i)\}}$, where $a(i)$ is the mean distance from the given observation to other elements of the same cluster, and $b(i)$ is the minimum, over the other $k-1$ clusters, of the mean distance from the $i$th observation to elements of the cluster. When $b(i)>a(i)$, as is usually but not always the case, the silhouette reduces to $1-a(i)/b(i)$, an intuitive measure of how clearly the $i$th observation belongs to its assigned cluster. An obvious CT extension is to define, for $t\in\cI$, $s(t)=\frac{b(t)-a(t)}{\max\{a(t),b(t)\}}$, where the mean distances $a(t),b(t)$ are defined as above but as integrals over clusters, divided by the length of the relevant cluster. It is straightforward to approximate these integrals by the trapezoidal rule, and thereby obtain $s(t)$, at each of a grid of equally spaced time points spanning $\cI$. The left panel of \figref{silfig} displays $s(t)$ for the CT 3-means clustering of the Chicago air pollution data of \secref{chisec}. Within most of the subintervals into which each cluster is divided, the silhouette width is low near the left boundary (as one would expect near a transition between clusters), rises to a peak, then falls again near the right boundary. For the long 1991--92 subinterval of cluster 3 discussed in the main text, there are two peaks with a  trough in between. This trough occurs around December 1991 and represents times that are on the borderline between the assigned cluster (3) and the cluster to which the winter is ordinarily assigned (1).

The right panel of \figref{silfig} shows the average of $s(t)$, over 6000 grid points, for CT $k$-means clustering of the air pollution data with $k$ ranging from 2 to 9. \cite{rousseeuw1987} suggests that ``one way to choose $k$ `appropriately' is to select that value of $k$ for which $\bar{s}(k)$ [mean silhouette width] is as large as possible.'' According to this widely adopted proposal, the plot indicates that $k=3$, while preferable to $k-4,\ldots,9$, gives a less successful clustering of this data set than $k=2$ (due, apparently, to the relatively low $s(t)$ within the spring/autumn cluster). On the other hand, \cite{rousseeuw1987} did not advocate a rigid policy of treating the mean-silhouette-maximizing $k$ as the true value to the exclusion of all others; he left open the possibility that different numbers of clusters might provide alternative informative views of the data. This is implied both by the scare quotes in the above quotation, and the fact that \cite{rousseeuw1987} himself presented both 2- and 3-cluster partitions of an example data set from political science. Adopting a similarly flexible approach, we felt that notwithstanding the mean silhouette results, $k=3$ gives a more interesting cluster partition than $k=2$ for the air pollution data, and we therefore chose to present the 3-means clustering in the text.

The 3-means clustering of this data set was run 50 times. In 46 instances, there were 54 cluster transitions, exactly as in  the solution described in \secref{chisec}. The timings of each of these transitions never differed by more than one day between any two of these 46 clusterings. In the other 4 instances, there were 40 cluster transitions, whose timings were again very consistent, with discrepancies that were uniformly less than one day. The 54-transition solution that was usually chosen had lower objective function (total within-cluster integrated squared distance) than the 40-transition solution; the ratio of between to total integrated squared distance, a metric that is analogous to a standard measure of clustering quality, is about 88\% for the former solution and about 80\% for the latter. These results suggest that for this application, 3-means clustering is quite stable, with about 90\% of starting values leading to an optimal partition. At the same time, the suboptimal solution attained in several instances underscores the importance of trying multiple starting values.

\section{Multivariate Gaussian processes}\label{simdet}
The true functions $\bx(\cdot)$ in the simulation study of \secref{simsec} arise from a multivariate Gaussian process,  which can be defined as follows \citep{chen2020}. Given a mean function  $\bmu:\cI\rightarrow\mathbb{R}^p$, within-curve covariance function $\Gamma:\cI\times\cI\rightarrow\mathbb{R}$ and $p\times p$ between-curve covariance matrix $\bSigma$, we say that $\bx:\cI\rightarrow\mathbb{R}^p$ arises from the multivariate Gaussian process MGP$(\bmu,\Gamma,\bSigma)$ if, for any $t_1,\ldots,t_n\in\cI$, the $n\times p$ matrix $\bX_{t_1,\ldots,t_n}\equiv[x_u(t_i)]_{1\leq i\leq n,1\leq u\leq p}$ has the matrix-variate normal distribution \citep{dawid1981, gupta1999} with $n\times p$ mean matrix having $i$th row $\bmu(t_i)^T$, between-row covariance matrix $\bGamma_{t_1,\ldots,t_n}\equiv[\Gamma(t_i,t_j)]_{1\leq i,j\leq n}$ and between-column covariance matrix $\bSigma$. 

The above process is unidentifiable in the sense that it is equal to MGP$(\bmu,h\Gamma,h^{-1}\bSigma)$ for any $h>0$. But if we assume stationarity,  identifiability can be established by letting $\Gamma$ be an autocorrelation function $\Gamma(s,t)=a(|s-t|)$ with $a(0)=1$, e.g., the squared exponential \eqref{squexp}.

 \section{Additional simulations}\label{moresim}
\begin{figure}
	\centering
	\includegraphics[width=\textwidth]{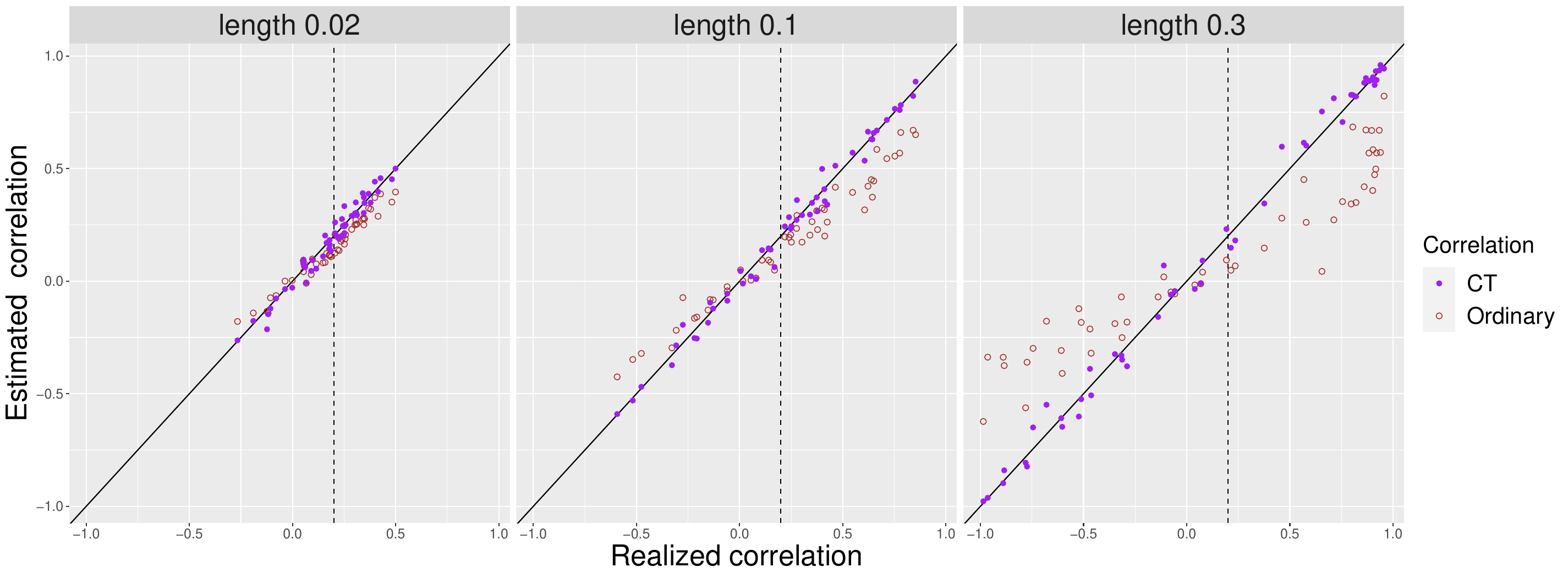}     
	\includegraphics[width=\textwidth]{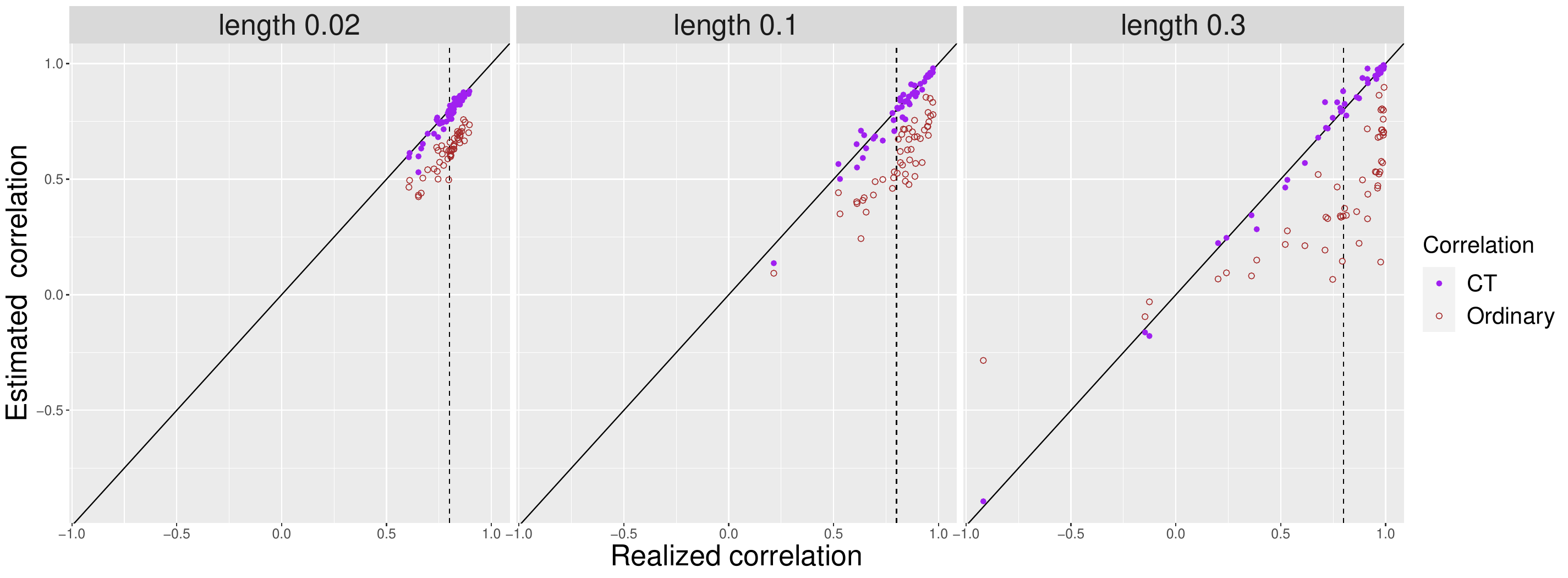}     
	\caption{Simulation results, analogous to those in the bottom row of \figref{biplabfig}, but with underlying between-curve correlation 0.2 (above) and 0.8 (below).}\label{biplabfig2}
\end{figure}

We conducted further simulations that were identical to those in \secref{simsec} except for the between-curve correlation. More specifically, as in \secref{simsec}, bivariate Gaussian processes as in \eqref{squexp}, with $\ell=0.02, 0.1, 0.3$, were observed at $n=500$ points with noise having standard deviation $\sigma=0.5$; but instead of between-curve correlation $\rho=0.5$, we generated 50 replicates with $\rho=0.2$ and 50 replicates with $\rho=0.8$.  The results, analogous to those in the bottom row of \figref{biplabfig}, are displayed in \figref{biplabfig2}. As in \secref{simsec}, the ordinary correlation estimates are attenuated and the CT correlations are much more accurate (points are closer to the line of identity).

\begin{figure}
	\centering
	\includegraphics[width=1.06\textwidth]{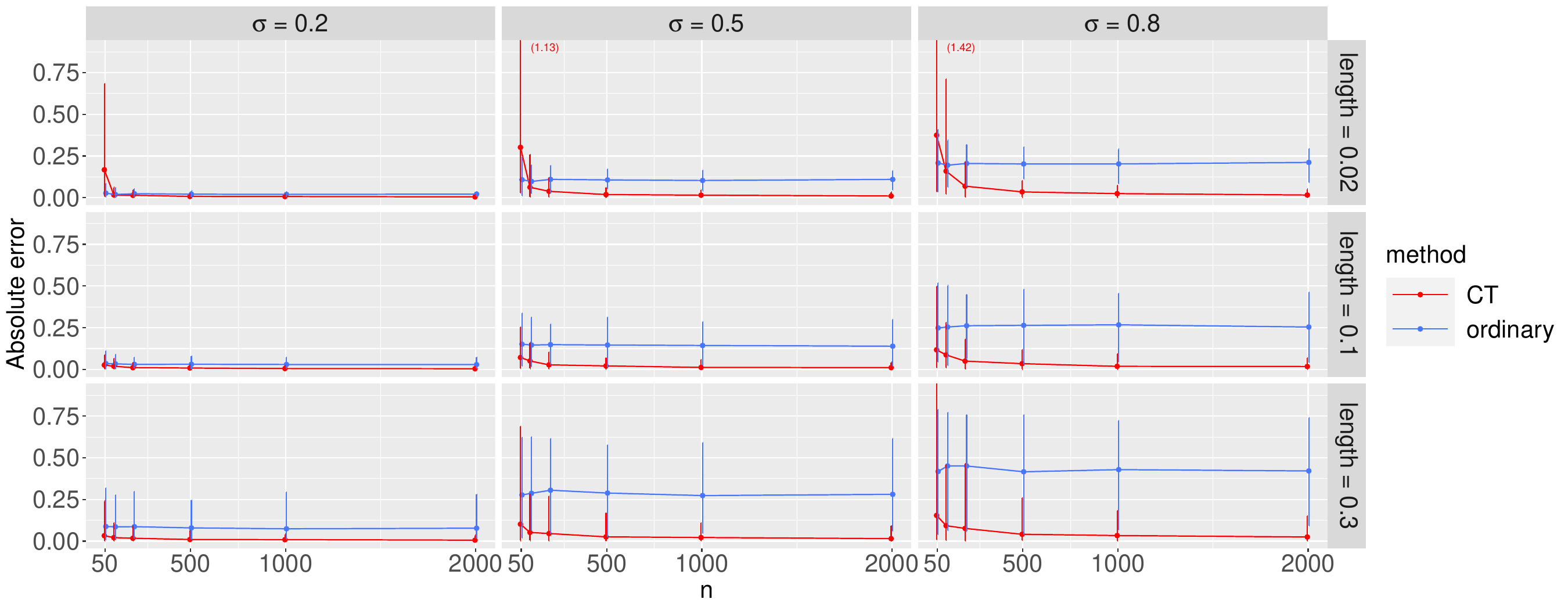}     
	\includegraphics[width=\textwidth]{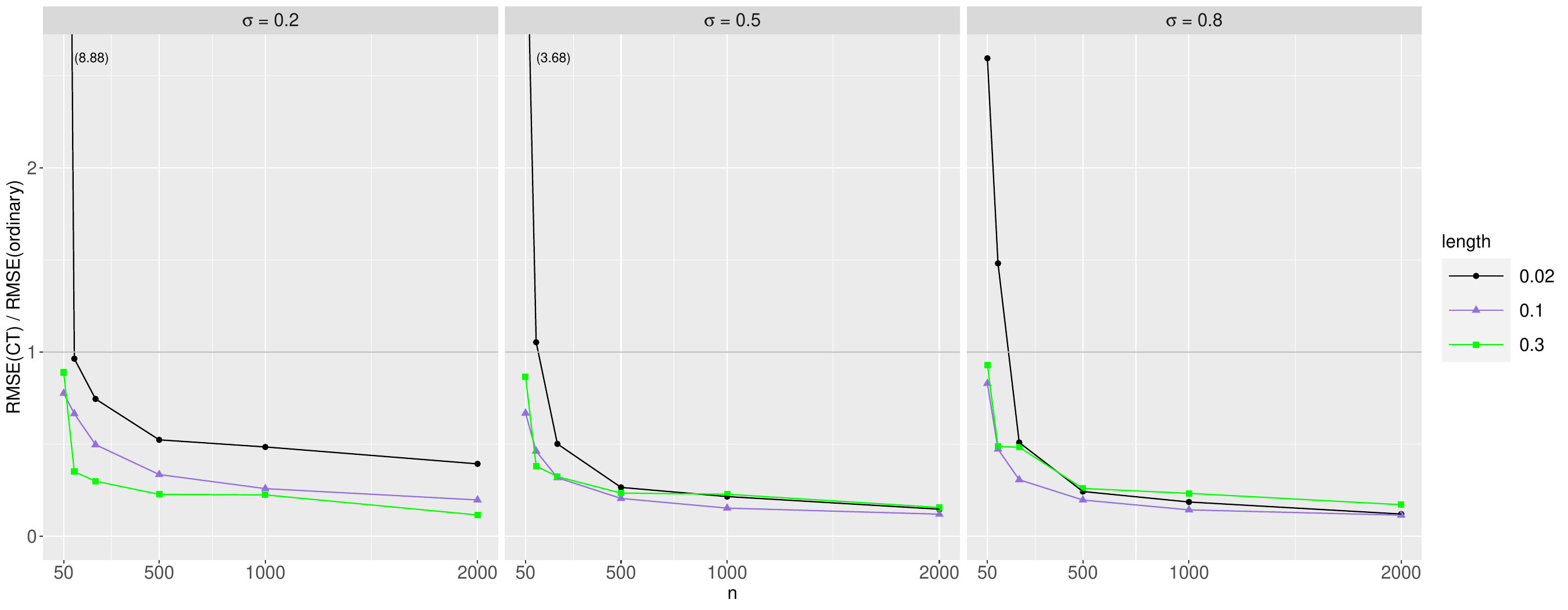}     
	\caption{Above: Median absolute error in estimating the between-curve correlation $r^*$ by ordinary and CT correlation, with error bars for 5th and 95th percentiles. Horizontal axis represents the number $n$ of points at which the curves are observed with noise, and $\sigma$ is the standard error of the noise. ``Length'' denotes the parameter $\ell$ in \eqref{squexp}. Below: Ratio of root mean square error in estimating $r^*$ by CT versus ordinary correlation.
		Several values lie beyond the vertical axis range, and are indicated in parentheses.}\label{biplabfig3}
\end{figure}

These results, along with those of \figref{biplabfig}, provide strong visual evidence of the superiority of CT over ordinary correlation when $n=500$ and $\sigma=0.5$. We performed a more extensive simulation study, with $\rho=0.5$ and $\ell=0.02, 0.1, 0.3$ as in \figref{biplabfig}, but now with 50 replicates for each combination of
$n=50,100,200,500,1000,2000$ and $\sigma=0.2, 0.5, 0.8$. The upper part of \figref{biplabfig3} shows the median of the absolute error in estimating the correlation $r^*$ between the two functions, i.e., $|\hat{r}-r^*|$ for ordinary correlation and $|\hat{r}^*-r^*|$ for CT correlation, along with error bars from the 5th to the 95th percentiles, for each setting. The performance of ordinary correlation worsens as $\sigma$ increases, and does not depend noticeably on $n$. CT correlation, on the other hand, performs poorly in most settings with $n=50$, but as $n$ increases its median error drops well below that for ordinary correlation. The lower part of \figref{biplabfig3}  shows the ratio of the root mean square error (RMSE) for CT versus ordinary correlation. For $\ell=0.02$, regardless of $\sigma$, CT correlation performs much worse with $n=50$ and somewhat worse with $n=100$ in terms of RMSE, but outperforms ordinary correlation for $n\geq 200$. For $\ell=0.1,0.3$, i.e.\ for smoother curves, CT correlation has lower RMSE for all values of $n$. 
In summary, CT correlation is more accurate than ordinary correlation except when the function is too bumpy and/or is observed too sparsely.

Computing time for the CT correlation estimate $\hat{r}^*$, including $B$-spline smoothing of the two functions, depended primarily on $n$. On an HP Pavilion x360 with 8 GB of RAM, with the other settings fixed at $\ell=0.02, \sigma=0.5$, the computing time in seconds was 0.35, 0.53, 0.97, 1.42 for $n=100,500,1000,2000$, respectively.

\bibliographystyle{chicago}

\bibliography{beta}

\end{document}

%% file: bold.tex
\newcommand{\ba}{\mbox{\boldmath $a$}}
\newcommand{\bb}{\mbox{\boldmath $b$}}
\newcommand{\bc}{\mbox{\boldmath $c$}}

\newcommand{\be}{\mbox{\boldmath $e$}}

\newcommand{\bm}{\mbox{\boldmath $m$}}

\newcommand{\bv}{\mbox{\boldmath $v$}}
\newcommand{\bw}{\mbox{\boldmath $w$}}
\newcommand{\bx}{\mbox{\boldmath $x$}}
\newcommand{\by}{\mbox{\boldmath $y$}}

\newcommand{\bB}{\mbox{\boldmath $B$}}
\newcommand{\bC}{\mbox{\boldmath $C$}}
\newcommand{\bD}{\mbox{\boldmath $D$}}

\newcommand{\bG}{\mbox{\boldmath $G$}}

\newcommand{\bI}{\mbox{\boldmath $I$}}

\newcommand{\bL}{\mbox{\boldmath $L$}}
\newcommand{\bM}{\mbox{\boldmath $M$}}

\newcommand{\bQ}{\mbox{\boldmath $Q$}}
\newcommand{\bR}{\mbox{\boldmath $R$}}
\newcommand{\bS}{\mbox{\boldmath $S$}}
\newcommand{\bT}{\mbox{\boldmath $T$}}

\newcommand{\bW}{\mbox{\boldmath $W$}}
\newcommand{\bX}{\mbox{\boldmath $X$}}
\newcommand{\bY}{\mbox{\boldmath $Y$}}

\newcommand{\bone}{\mbox{\bf 1}}

\newcommand{\bGamma}{\mbox{\boldmath $\Gamma$}}

\newcommand{\bmu}{\mbox{\boldmath $\mu$}}

\newcommand{\bSigma}{\mbox{\boldmath $\Sigma$}}

\newcommand{\bphi}{\mbox{\boldmath $\phi$}}
\newcommand{\bPhi}{\mbox{\boldmath $\Phi$}}

\newcommand{\cC}{\mathcal{C}}

\newcommand{\cI}{\mathcal{I}}

\newcommand{\cS}{\mathcal{S}}

\newcommand{\cX}{\mathcal{X}}